\documentclass[11pt,letterpaper]{article}
\usepackage[top=1in,bottom=1in,left=1in,right=1in]{geometry}

\usepackage{amsmath,amssymb}
\usepackage{changepage}
\usepackage{setspace,newfloat,caption}
\usepackage{mathtools}
\usepackage{graphicx,tabularx}
\usepackage{dcolumn}
\usepackage{bm}
\usepackage[export]{adjustbox}

\usepackage{mathtools}
\mathtoolsset{showonlyrefs}

\usepackage{sectsty}
\subsubsectionfont{\itshape}

\usepackage[utf8]{inputenc}

\usepackage{textcomp,marvosym}

\usepackage[hidelinks]{hyperref}
\usepackage{nameref}

\usepackage{microtype}
\DisableLigatures[f]{encoding = *, family = * }

\usepackage[table]{xcolor}

\usepackage{array}

\newcolumntype{+}{!{\vrule width 2pt}}

\newlength\savedwidth

\newcommand{\beginSupplement}{%
       \setcounter{section}{0}
       \renewcommand{\thesection}{S\arabic{section}}%
        \setcounter{table}{0}
        \renewcommand{\tablename}{{Table}}  
        \renewcommand{\thetable}{{S\arabic{table}}}%
        \setcounter{figure}{0}
        \renewcommand{\figurename}{{Figure}}    
        \renewcommand{\thefigure}{{S\arabic{figure}}}%
     }   

\usepackage[aboveskip=12pt,labelfont=bf,labelsep=period,justification=raggedright,singlelinecheck=off]{caption}
\renewcommand{\figurename}{Figure}

\usepackage{natbib}

\makeatletter
\renewcommand{\@biblabel}[1]{\quad#1.}
\makeatother

\date{}

\usepackage{lastpage,fancyhdr,graphicx}
\usepackage{epstopdf}
\usepackage{caption}
\usepackage[T1]{fontenc}
\usepackage{lmodern}
\usepackage{epsfig, graphicx,setspace}
\usepackage{float}
\usepackage{latexsym}
\usepackage{amstext}
\usepackage{amssymb}
\usepackage{graphics}
\usepackage{color}
\usepackage{multirow}

\def\fref#1{{Figure~\ref{#1}}}

\def\sref#1{{\emph{\nameref{#1}}}}

\renewcommand{\figurename}{\textbf{{Figure}}}

\begin{document} 

\begin{flushleft}

\begin{spacing}{1.5}
\textbf{\Large  The Key Parameters that Govern Translation Efficiency}
\end{spacing}

\vspace{5mm}
Dan D. Erdmann-Pham\textsuperscript{1}, 
Khanh Dao Duc\textsuperscript{2} and
Yun S. Song\textsuperscript{2,3,4,*}
\\
\bigskip
\textbf{1} Department of Mathematics, University of California, Berkeley, CA 94720, USA
\\
\textbf{2} Computer Science Division, University of California, Berkeley, CA 94720, USA
\\
\textbf{3} Department of Statistics, University of California, Berkeley, CA 94720, USA
\\
\textbf{4} Chan Zuckerberg Biohub, San Francisco, CA 94158, USA
\bigskip

* Lead Contact and Corresponding Author: yss@berkeley.edu
\end{flushleft}

\vspace{5mm}

\subsection*{ABSTRACT}
Translation of mRNA into protein is a fundamental yet complex biological process with multiple factors that can potentially affect its efficiency.  Here, we study a stochastic model describing the traffic flow of ribosomes along the mRNA (namely, the inhomogeneous $\ell$-TASEP), and identify the key parameters that govern the overall rate of protein synthesis, sensitivity to initiation rate changes, and efficiency of ribosome usage.  By analyzing a continuum limit of the model, we obtain closed-form expressions for stationary currents and ribosomal densities, which agree well with Monte Carlo simulations.  Furthermore, we completely characterize the phase transitions in the system, and by applying our theoretical results, we formulate design principles that detail how to tune the key parameters we identified to optimize translation efficiency.  Using ribosome profiling data from \emph{S. cerevisiae}, we shows that its translation system is generally consistent with these principles.  Our theoretical results have implications for evolutionary biology, as well as synthetic biology.

\vspace{2cm}
\noindent
To appear in \emph{Cell Systems}: 
\url{https://doi.org/10.1016/j.cels.2019.12.003}\\
Copyright: CC BY-NC-ND license
\newpage

\section*{INTRODUCTION}

Being a major determinant of gene expression and protein abundance levels  \citep{lu2007absolute, kristensen2013protein}, translation of mRNA into polypeptides is one of the most fundamental biological processes underlying life.
The extent to which this process is regulated and shaped by the sequence landscape has been widely studied over the past decades \citep{dever2016mechanism, hanson2018codon, quax2015codon}, revealing many intricate mechanisms that may affect translation dynamics. 
From a more global perspective, however, it has been challenging to integrate these findings to elucidate the key factors that govern translation efficiency.
Indeed, translation is a complex process that depends on many parameters, including the initiation rate, site-specific elongation rates (which can vary substantially along a given transcript), and the termination rate.
How does the overall rate of protein synthesis depend on these parameters? 
To make the problem more concrete, suppose that the goal is to achieve the fastest rate of protein production while minimizing the cost.  Would choosing the ``fastest'' synonymous codon at each site do the job?
If the local elongation rate changes at a particular site, would it necessarily affect the overall rate of protein synthesis?  If not, then which parameters actually matter?
Aside from achieving a desired protein production rate, how does a translation system make efficient use of available resources, particularly the ribosomes?
These are important questions in molecular and evolutionary biology, as well as synthetic biology, but challenging to answer because there are many parameters involved -- for a transcript consisting of $N$ codons, one has to analyze a model with about $N$ parameters, which is seemingly intractable when $N$ is large.

In this article, we develop a theoretical tool to answer the above questions.
Our work hinges on analyzing a mathematical model that describes the traffic flow of ribosomes, which mediate translation by moving along the mRNA transcript.
Beginning with 
\citet{macdonald1968}, most mechanistic
  studies of translation dynamics have been based on the so-called Totally
Asymmetric Simple Exclusion Process (TASEP), a probabilistic model that explicitly describes
the flow of particles along a lattice \citep{zia2011modeling,zur2016predictive}. As a classical model of
 transport phenomena in non-equilibrium, the TASEP has attracted wide interest from mathematicians and
physicists \citep{blythe2007nonequilibrium}.  To describe translation realistically, however,
a generalized version of the model needs to be employed, taking into account the extended size of the ribosome and the heterogeneity of the elongation rate along the transcript. Under such general conditions, critical questions have hitherto remained open; 
in particular, identifying the parameters most crucial to 
the current and particle density has proven elusive.

Here we carry out a theoretical analysis of a generalized version of the TASEP
and obtain analytic results that provide practical insights into translation dynamics.  Our approach is to study the process in a continuum limit called the hydrodynamic limit, which leads to a general PDE satisfied by the density of particles. Upon solving this PDE,
we obtain exact closed-form expressions for stationary currents and particle densities that agree very well with Monte Carlo simulations of the original TASEP model.
Furthermore, we provide a complete  characterization of phase transitions in the system.
These results allow us to identify the key parameters that govern translation dynamics, and to
formulate a set of specific design principles for optimizing translation efficiency in terms of protein production rate and resource usage.  Using experimental ribosome profiling data of \textit{S. cerevisiae}, we show that the translation system of this organism is generally efficient according to the design principles we found.

\section*{RESULTS}

We first present our theoretical results on a mathematical model of translation and identify the
key parameters that govern its dynamics.  We then apply our theoretical results to formulate four simple design principles that detail how to tune these parameters to optimize the overall rate of protein synthesis and efficiency of ribosome usage.  We then analyze ribosome profiling data of \emph{S. cerevisiae} and demonstrate that its translation system is generally efficient, consistent with the design principles we found.

\subsection*{Theoretical Results on a Stochastic Model of Translation}

\subsubsection*{Model description of the inhomogeneous $\ell$-TASEP}
\begin{figure}[h]  
    \centering
    \includegraphics[width=\linewidth]{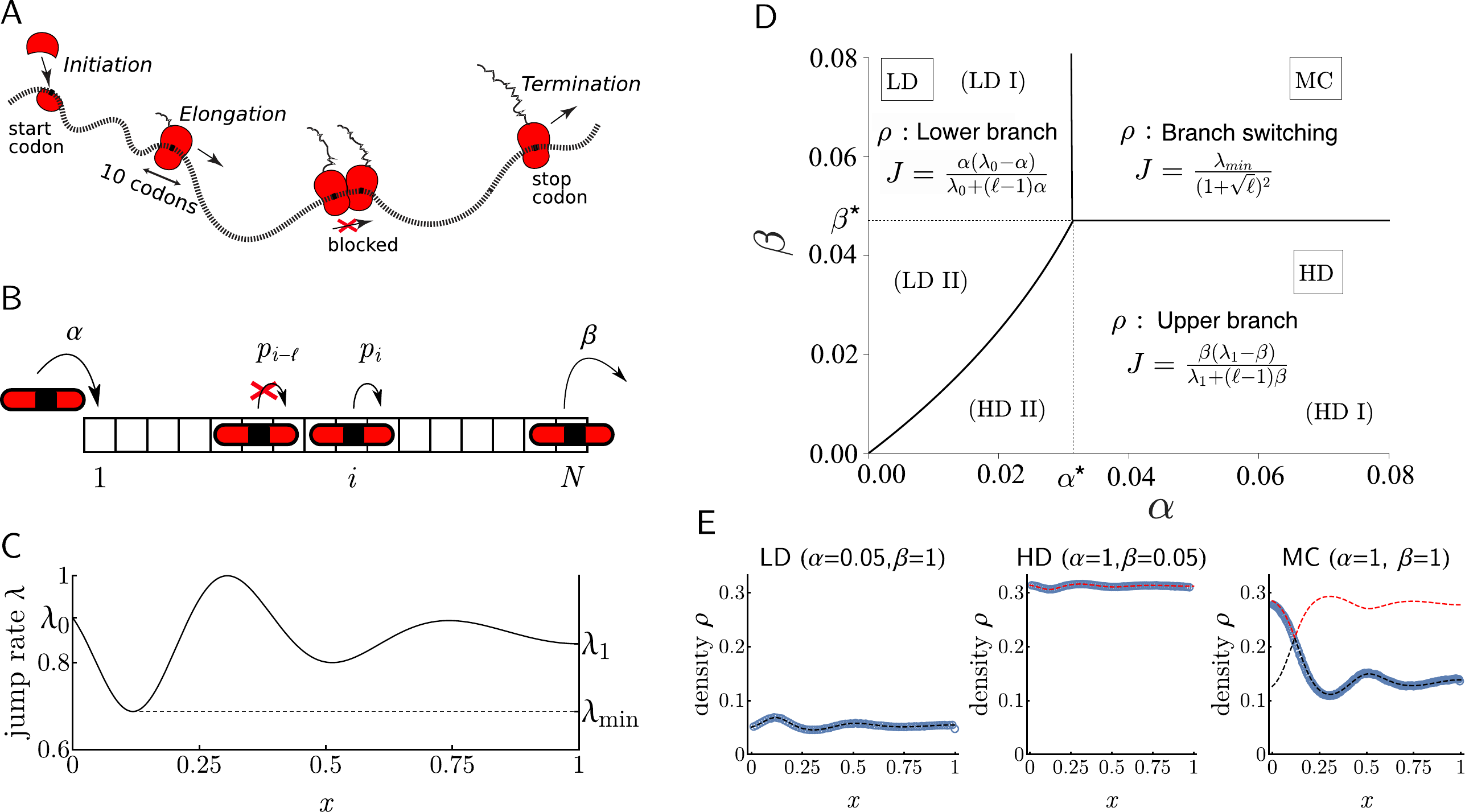}
    \caption{\textbf{Illustration of the translation process, the inhomogeneous 
        $\ell$-TASEP with open boundaries, and its phase diagram.} 
        \textbf{A:} Ribosomes initiate translation at the mRNA $5'$ end, elongate
        the polypeptide by decoding one codon at a time, and eventually terminate the 
        process by detaching from the transcript.  \textbf{B:} Particles 
        (of size $\ell=3$ here) enter the lattice at rate $\alpha$ and a particle at 
        position $i$ (here defined by the position of the midpoint of the particle) 
        moves one site to the right at rate $p_i$, provided that the move is not blocked 
        by another particle in front. \textbf{C:} Example rate function with
        key parameters shown. \textbf{D:} The phase diagram is completely 
        determined by $\lambda_0, \lambda_1, \lambda_{\min}$ and $\ell$. In this example,
        $(\lambda_0, \lambda_1, \lambda_{\min}, \ell) = (0.9,0.3,0.1,10)$. All phase
        transitions are continuous in $J$ and, unless $\lambda_{\min}$ coincides 
        with $\lambda_0$ or $\lambda_1$, discontinuous in $\rho$.
        \textbf{E:} Simulated results for $\ell=3,N=800$, and $\lambda$ as in
        \textbf{C} are compared with theoretical predictions. Dashed black and red
        lines represent upper and lower branches of solutions to
        \eqref{eq:hydro_limit}. Circles are averaged counts over $5\times 10^7$
        Monte-Carlo steps after $10^7$ burn-in cycles. 
    }
    \label{fig:illustration}
\end{figure}

At a high level, translation of mRNA involves three types of movement of the ribosome, as illustrated in \fref{fig:illustration}A: 
1) Initiation -- a small ribosomal subunit enters the open reading frame so that its A-site is positioned at the second codon and then a large ribosomal subunit binds with the small subunit.
2) Elongation -- the nascent peptide chain gets elongated by one amino acid and the ribosome moves forward by one codon.
3) Termination -- the ribosome with its A-site at the stop codon unbinds from the transcript. 
An important point to note is that more than one ribosome can translate the same mRNA transcript simultaneously, so the movement of a ribosome can be obstructed by another ribosome in front, similar to what happens in a traffic flow on a one-lane road.  Such interaction is what makes the dynamics difficult to analyze.

We model the flow of ribosomes on mRNA using a generalized TASEP, called the inhomogeneous $\ell$-TASEP, on a one-dimensional lattice with $N$ sites (see \fref{fig:illustration}B).
In this process, each particle (corresponding to a ribosome in mRNA translation) is of a fixed size $\ell\in \mathbb N$
and is assigned a common reference point (e.g., the midpoint in the example illustrated in 
\fref{fig:illustration}B). The position of a particle is defined as the location of its 
reference point on the lattice. A configuration of particles is denoted by the vector 
$\boldsymbol{\tau} = (\tau_1, \ldots, \tau_N)$, where $\tau_i = 1$ if the
$i^{\textnormal{th}}$ site is occupied by a particle reference point and $\tau_i = 0$ 
otherwise. The jump rate at site $i$ of the lattice is denoted by $p_i>0$.
During every infinitesimal time interval $dt$, each particle located at position
$i \in \left\{ 1, \ldots , N-1 \right\}$  has probability $p_i dt$ of jumping exactly one 
site to the right, provided that the next $\ell$ sites are empty; particles at positions 
between $N-\ell+1$ and $N$, inclusive, never get obstructed. Additionally, a new particle 
enters site~1 with probability $\alpha dt$ if $\tau_i =0$ for all $i=1,\ldots,\ell$.  
If $\tau_N =1$, the particle at site $N$ exits the lattice with probability $\beta dt$.  
The parameter $\alpha$ is called the entrance (or initiation) rate, while
$\beta$ is called the exit (or termination) rate.

\subsubsection*{The hydrodynamic limit}
The key quantities of interest are the stationary probability $\langle \tau_i
\rangle$ of any individual site $i$ being occupied or not, and the
current (or flux) $J$ of particles in the system. 
In the corresponding translation process, these quantities reflect the local ribosomal density and the
protein production rate, respectively.

In the special case of the homogeneous  $1$-TASEP ($p_i = p$ for all $i$ and $\ell = 1$), the stationary 
distribution of the process decomposes into matrix product states, 
which can be treated analytically \citep{derrida1993}.  Unfortunately, 
in the general case this approach is intractable, necessitating alternative methods 
such as the hydrodynamic limit.
When $\ell >1$, deriving the hydrodynamic limit is not straightforward, however,
as the process does not possess stationary product measures \citep{schonherr2004}. 
To tackle this problem, we mapped the $\ell$-TASEP to another interacting particle system
called the zero range process (ZRP, see \sref{sec:methods_hydro_limit} of STAR Methods and
Figure~S1), whose hydrodynamic limit, assuming it exists, can be 
derived from the 
associated master equation. More precisely, we obtained the hydrodynamic limit through 
Eulerian scaling of time and space by a factor $a = N^{-1}$, and by following its 
dynamics on scale $x$ such that $k = \left \lfloor{\frac{x}{a}}\right \rfloor$, 
for $1< k < N$ \citep{rezakhanlou1991hydrodynamic}.
Implementing this limiting procedure for the ZRP and mapping it back to the inhomogeneous 
$\ell$-TASEP, we found that the limiting occupation density 
$\rho(x,t) := \mathbb{P}(\tau_{k}(t)=1)$, assuming its existence, satisfies the nonlinear PDE
\begin{equation}
\partial_t \rho = -\partial_x\left[\lambda(x) \rho\, G(\rho) \right] +
            \frac{a}{2} \partial_{xx} \left[ \lambda(x)G(\rho) \right] + O(a^2), \hspace{3mm} \label{eq:hydro_limit}  
\end{equation}
where $G(\rho) = \dfrac{1- \ell \rho}{1- (\ell-1)\rho}$ and $\lambda$ is a 
differentiable extension of $(p_1, \ldots, p_N)$, such that 
$\lambda(x) = \lambda(ka) = p_k $. More generally, this PDE takes the form of a 
conservation law with systematic and diffusive currents $J$ and $J_D$, given by
        \begin{align}
            J(\rho, x) = \lambda(x)\rho G(\rho) \hspace{5mm}\text{and}\hspace{5mm}
            J_D(\rho, x) = \dfrac{\lambda(x) \rho }{ 1-(\ell-1)\rho }.
            \label{eq:currents}
        \end{align}
As $a \ll 1$, the systematic current dominates and solutions of
\eqref{eq:hydro_limit} generically converge locally uniformly on $(0,1)$ to so-called 
entropy solutions of
\begin{equation}
    \partial_t \rho = - \partial_x\left[\lambda(x) \rho\, G(\rho) \right].
    \label{eq:first_order}
\end{equation}
Further details and relevant calculations are provided in
\sref{sec:methods_hydro_limit} of STAR Methods.

\subsubsection*{Particle densities, currents and phase transitions}
\label{sec:particle_densities}
The first order nonlinear PDE given by \eqref{eq:first_order} can be solved using the method
of characteristics
\citep{evans2010partial}, which describes the evolution of differently dense ``patches'' 
of particles over time. Solving for the characteristics yields two branches of solutions,
which we call ``upper'' and ``lower'' branches, while the boundary conditions imposed by 
$\alpha$ and $\beta$ determine which branch is taken by the stationary density of particles 
(see \sref{sec:phase_transition} of STAR Methods).
As a consequence, the behavior of the system is characterized by a phase diagram in 
$\alpha$ and $\beta$. Moreover, this phase diagram depends on only few parameters of 
the system (see \fref{fig:illustration}C): the size of particles $\ell$, the
jump rates at the boundaries, $\lambda_0:= \lambda(0)$ and $\lambda_1:=
\lambda(1)$, and the minimum jump rate $\lambda_{\min}:= \min 
\{ \lambda(x) : x\in [0,1]\}$. In particular, these parameters determine the
critical initiation and termination rates, $\alpha^*$ and $\beta^*$, that are associated 
with phase transitions.  More precisely, the critical initiation rate $\alpha^*$ is given by
\begin{equation}
    \alpha^* = \frac{\lambda_0 - (\ell-1)J_{\max}}{2} \! \left[ 1 - \sqrt{
    1 - \dfrac{4 \lambda_0 J_{\max}}{\left[\lambda_0 - (\ell-1)J_{\max}\right]^2}}\, \right],
    \label{eq:alphastar}       
\end{equation}
where $J_{\max}=\frac{\lambda_{\min}}{(1+\sqrt{\ell})^2}$.
Note that $\alpha^*$ is determined by the jump rates $\lambda_0$ and $\lambda_{\min}$.
In the context of translation dynamics, this means that $\alpha^*$ will be specific to each gene, as different genes will likely have different values of $\lambda_0$ and $\lambda_{\min}$.
For a fixed $\lambda_0$, the critical rate $\alpha^*$ increases as $\lambda_{\min}$ increases.  
For a fixed $\lambda_{\min}$, it turns out that $\alpha^*$ satisfies
\begin{equation}
 \frac{\lambda_{\min}}{(1+\sqrt{\ell})^2} \leq \alpha^* \leq  \frac{\lambda_{\min}}{1+\sqrt{\ell}},
 \label{eq:alphastar_range}
\end{equation}
where the lower bound is achieved as $\lambda_0\to\infty$, while the upper bound is achieved when $\lambda_0=\lambda_{\min}$.  More generally, for a fixed $\lambda_{\min}$, the critical initiation rate $\alpha^*$ decreases as $\lambda_0$ increases.
The critical termination rate $\beta^*$ is obtained from \eqref{eq:alphastar} by replacing $\lambda_0$ with $\lambda_1$.  Hence, for mRNA translation, $\beta^*$ is also gene-specific, determined by the key elongation rates $\lambda_1$ and $\lambda_{\min}$.

The resulting phase diagram, which generalizes previous formulas for the
homogeneous 1-TASEP \citep{derrida1993}, is summarized as follows 
(see \fref{fig:illustration}D):

\medskip
1. If $\alpha < \alpha^{\ast}$ and $\beta > \beta^{\ast}$ (LD I): In this regime 
    the flux is limited by the initiation rate, leading to a \textit{low
    density} profile. The corresponding current assumed by the system is
\begin{equation}
   J_L = \frac{\alpha(\lambda_0 - \alpha)}{\lambda_0 + (\ell-1)\alpha}, \label{eq:J_L}\\
\end{equation}
while the site-specific particle density is
\begin{equation}
\rho_L (x) =  \frac{1}{2\ell} + \frac{J_L(\ell-1)}{2\ell \lambda(x)} - \sqrt{
            \left[\frac{1}{2\ell} +\frac{J_L(\ell-1)}{2\ell \lambda(x)} \right]^2 -
        \dfrac{J_L}{\ell\lambda(x)} }. \label{eq:rho_l}
\end{equation}

2.  If $\alpha > \alpha^{\ast}$ and $\beta < \beta^{\ast}$ (HD I): Now the flux is 
    limited by the particle exit rate, resulting in a \textit{high density} regime. 
    The associated current $J_R$ and density $\rho_R$ are identical to $J_L$ 
    (\eqref{eq:J_L}) and $\rho_L$ (\eqref{eq:rho_l}), respectively, with $\lambda_0$ 
    and $\alpha$ replaced by $\lambda_1$ and $\beta$.

3. If $\alpha < \alpha^{\ast}$ and $\beta < \beta^{\ast}$ (LD II and HD II): The
    steady state is determined by the sign of $J_L - J_R$ (computed as above).
    If it is positive ($ J_L > J_R$), the system is in a low density regime  with current and density given by $J_L$ and $\rho_L$, respectively. Conversely, if it is negative, 
      the system is in a high density regime with $J_R$ and $\rho_R$ as the current and density.
      
4. If $\alpha > \alpha^{\ast}$ and $\beta > \beta^{\ast}$ (MC):
    The system carries the \emph{maximum possible current }(also referred to as the 
    transport capacity of the system) 
    \begin{equation}
        J_{\max} = \dfrac{\lambda_{\mathrm{min}}}{( 1+\sqrt{\ell} )^2},
        \label{eq:maximum_current}
    \end{equation}    
    which is limited only by the minimum elongation rate $\lambda_{\min}$. Its density
    is characterized by qualitatively different profiles to the left and right of
    $x_{\mathrm{min}} = \arg\min_x \lambda(x)$:  For $x< x_{\mathrm{min}}$,
    $\rho(x)$ is described by the upper branch (obtained by replacing
    $J_R$ with $J_{\max}$ in the equation for $\rho_R$), 
    while for $x> x_{\mathrm{min}}$, $\rho(x)$ is described by the lower branch (obtained by
    replacing $J_L$ with $J_{\max}$ in $\rho_L$). That is, a branch switch
    occurs at $x_{\min}$ (where $\rho(x_{\min}) = (1+\sqrt{\ell})^{-2}$). We proved more
    generally that every global minimum of $\lambda$ regulates the traffic of particles 
    (like a toll reducing the traffic flow) in this fashion: incoming densities to the 
    left of it are always described by the upper branch whereas outgoing particles on 
    the right follow the lower branch. In particular, this implies that in the case 
    of multiple global minima, the density between two consecutive minima must undergo 
    a discontinuous jump from lower to upper branch (for more details, see
    \sref{sec:phase_transition} of STAR Methods and Figure~S2).

\subsubsection*{Novel phenomena and applicability to discrete lattices}
\begin{figure}[p]
    \centering
    \includegraphics[width=.9\linewidth]{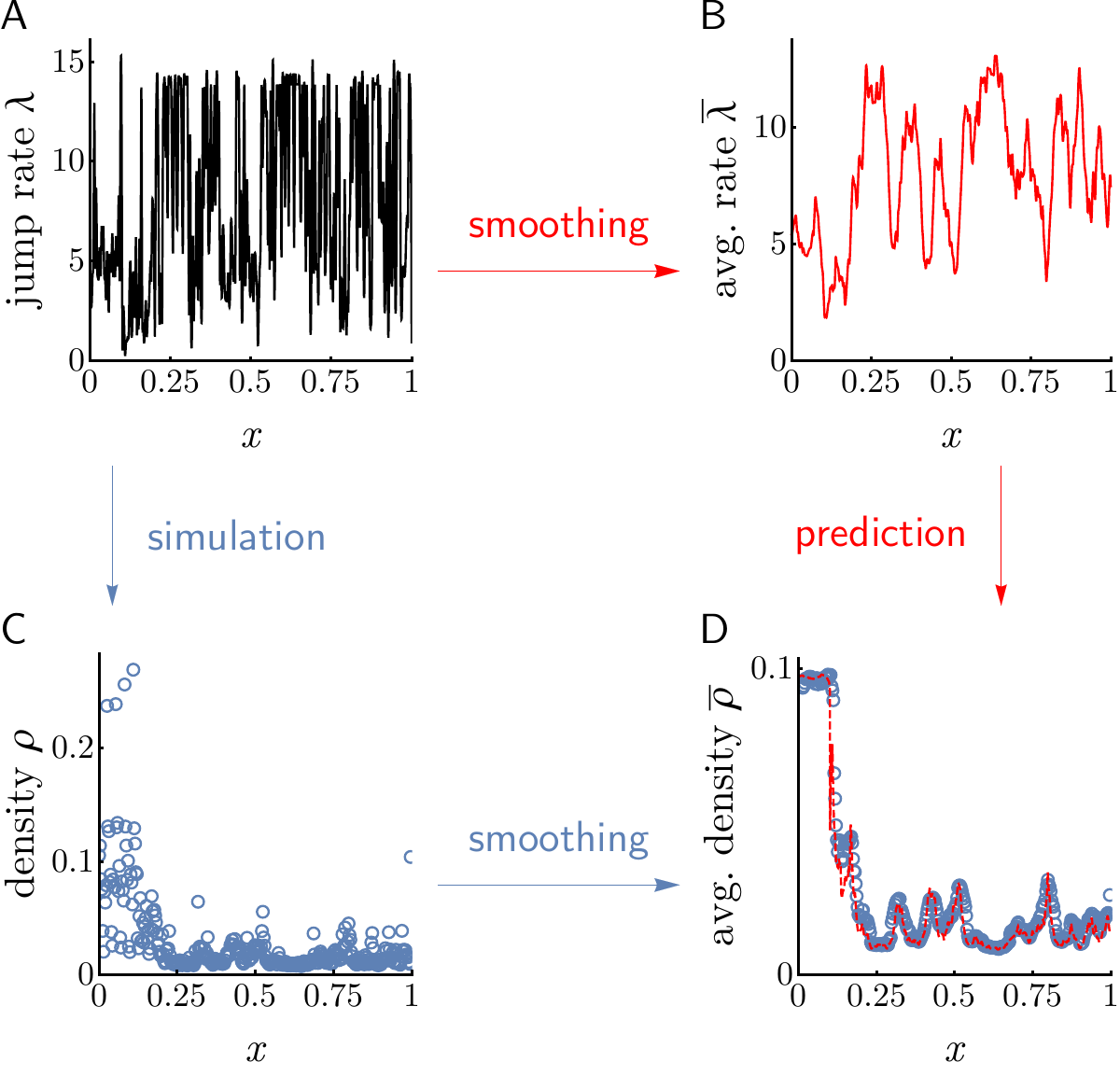}
    \caption{\textbf{Local averaging reproduces hydrodynamic limit in lattices
            with discontinuous rate functions.} Applying the hydrodynamic
            theory to smoothed jump rates correctly predicts smoothed density
            profiles and currents. \textbf{A:} Elongation rates of the yeast
            gene YHR025W arbitrarily chosen from \citet{daoduc2018impact} (see
            \sref{sec:yeast} for further details).
            \textbf{B:} Smoothed elongation rates obtained by applying a
            ten-codon moving average to the raw profile in \textbf{A}.
            \textbf{C:} Density profile resulting from simulation (as in
            \fref{fig:illustration}E except with $\ell = 10, N = 357$) 
            under discontinuous profile in \textbf{A}.
            \textbf{D:} The hydrodynamic density profile (dashed red)
            associated with the smoothed elongation rates of \textbf{B} reproduces the
            smoothed density profile obtained from averaging the raw densities
            in \textbf{C}
            by a moving ten-codon window. Similarly, simulated and predicted currents are in
            excellent agreement ($0.1072$ and $0.1077$, respectively).
        }
    \label{fig:smoothing}
\end{figure}

As shown in \fref{fig:illustration}E, for smooth rate functions the densities predicted by 
our analysis agree well with Monte Carlo simulations in all regimes of the phase diagram.
In the context of translation dynamics, however, elongation rates
are typically less regular, exhibiting substantial fluctuations throughout the
entire transcript (see \fref{fig:smoothing}A). Despite this lack of regularity,
the hydrodynamic limit can still be employed to describe local averages of such
a system. In particular, smoothing particle profiles by windows of length $\ell$
reproduces parameters that closely match hydrodynamic predictions
(see \sref{sec:applicability} of STAR Methods and Figure~S3). Hence,
all subsequent analyses described below will pertain to elongation rate profiles
smoothed by a ten-codon moving average. A noteworthy consequence of the above
results is that local averages of elongation rates are more predictive of
overall translation dynamics than their non-smoothed counterparts. In
particular, the location at which branch switching occurs in the MC regime is governed by 
$x_{\min} = \arg\min_x \{\overline{p}_x\}/N$ which may be, and in many cases is,
considerably different from $\arg\min_x\{ p_x\}/N$ 
(cf. Figure~S3).

We highlight a few novel phenomena in our generalization of the homogeneous
$1$-TASEP: 
First, extending particles to size $\ell > 1$ and lowering the
limiting jump rate $\lambda_{\min}$
reduces both the transport capacity $J_{\max}$ 
and the critical rates ($\alpha^*$ and
$\beta^*$) for entrance and exit, 
leading to an enlarged MC phase region. This is
expected as fewer particles are needed to saturate the lattice, and distances
between particles are larger, which in turn limits the number of particles able to
cross a site per given time.  
This phenomenon is quantified precisely using our explicit
expressions for $\alpha^{\ast}, \beta^{\ast}$, and $J_{\max}$ (see
\eqref{eq:alphastar} and \eqref{eq:maximum_current}). 
Second, the inhomogeneity in $\lambda$ may
deform the LD-HD phase separation from being a straight line in the homogeneous 
$\ell$-TASEP \citep{chou2004clustered} to a generally nonlinear curve (see 
\fref{fig:illustration}D) determined by
solutions $(\alpha,\beta)$ of
\begin{equation}
\frac{\alpha(\lambda_0 - \alpha)}{\lambda_0 + (\ell-1)\alpha} = \frac{\beta(\lambda_1 - \beta)}{\lambda_1 + (\ell-1)\beta},
\label{eq:nonlinear_transition}
\end{equation}
corresponding to the condition $J_L=J_R$.
This is a consequence of $\alpha$ and
$\beta$ affecting the system at different scales whenever $\lambda_0 \neq
\lambda_1$, resulting in a phase diagram that is no 
longer symmetric.
Lastly, our observation of density profiles performing branch switching in the MC phase
was indiscernible in the homogeneous case, as the high density and low density 
branches merge into a single value (viz. $\rho = \frac{1}{\sqrt{\ell} + \ell} $).

\subsection*{Application: Design Principles for Translational Systems}

We sought to apply our theoretical analysis to understand how the translational system can be regulated and optimized with regard to protein synthesis rate and ribosome usage. The hydrodynamic theory developed above singles out the key parameters that determine
the current and particle densities. We illustrate in \fref{fig:sensitivity} how $\lambda_0$, $\lambda_{\min}$,  and $x_{\min}$ impact the current capacity, its sensitivity to the initiation rate $\alpha$, and the global particle density, suggesting the following principles:

\begin{figure}[p]
    \centering
    \includegraphics[width=.6\linewidth]{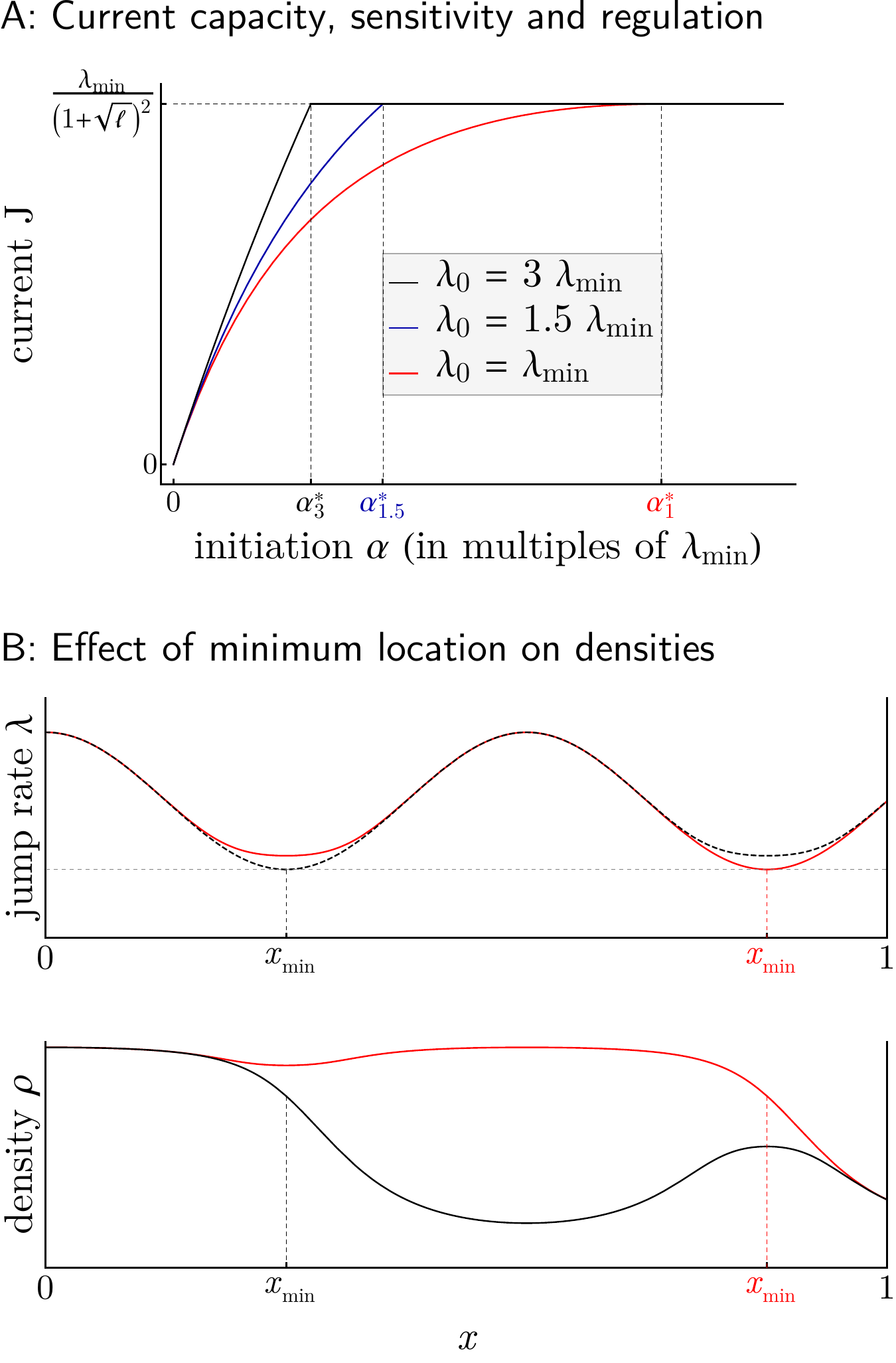}
    \caption{\textbf{Main determinants of current and particle densities.} 
        \textbf{A:} We plot the current $J$ in LD and MC against the
        initiation rate $\alpha$, for various choices of $\lambda_0$. While 
        $\lambda_{\min}$ governs the maximum current at which $J$ reaches a plateau 
        (coinciding with the transition
        from LD to MC), changing the size of $\lambda_0$ results in changes in 
        $\partial_{\alpha} J$, the sensitivity of $J$ with respect to $\alpha$.
        Distinct configurations of $\lambda_{\min}$ and $\lambda_0$ give rise to vastly
        different dependencies of $J$ on $\alpha$, suggesting 
        different responses to global changes in the ribosome pool. 
        $\alpha^*_3$, $\alpha^*_{1.5}$, and $\alpha^*_1$ correspond to the $\alpha^*$ value (in units of $\lambda_{\min}$) when $\lambda_0 = 3\lambda_{\min}, \lambda_0 =1.5\lambda_{\min},$ and  $\lambda_0 =\lambda_{\min}$, respectively.
        \textbf{B:} Two elongation rate profiles that differ slightly in overall shape,
        but drastically in their position $x_{\min}$ of minimum elongation are plotted (top
        panel) together with their associated MC ribosome densities (bottom panel).
        The branch switching phenomenon has extreme consequences for equilibrium particle 
        densities and hence ribosomal costs, with elongation rate profiles achieving minimum rates
        close to the initiation site (top, dotted black curve) benefiting from drastic 
        savings (bottom, black curve) compared to otherwise similar profiles (red curves).
}
    \label{fig:sensitivity}
\end{figure}

1. \textit{The initiation rate $\alpha$ (and not termination rate $\beta$) should regulate the production rate $J$.} 
        As shown by our analysis of the current, any value of the current that lies below the
        system's production capacity $J_{\max}$ can be attained through either HD or LD regime. In
        order to avoid overuse of resources, however, a transcript should always
        operate in LD, where the main determinant for currents is the initiation
        rate $\alpha$ (cf. \eqref{eq:J_L}). To guarantee LD profiles,
        termination rates merely need to exceed the critical
        value $\beta^{\ast}$, whereas initiation rates are more tightly
        controlled, varying between $0$ and $\alpha^{\ast}$.  Within this interval, the current $J$ increases with $\alpha$ according to \eqref{eq:J_L}, as illustrated in \fref{fig:sensitivity}A.
        
2. \textit{The minimum elongation rate $\lambda_{\min}$ determines the production
        capacity $J_{\max}$.}  As $\alpha$ increases in the LD regime, the current $J$ reaches a plateau that is associated with the maximal current (MC) regime (see \fref{fig:sensitivity}A). By \eqref{eq:maximum_current}, the maximum possible current
        is directly proportional to $\lambda_{\min}$, which therefore sets the range within 
        which production rates may
        vary. Large values of $\lambda_{\min}$ allow for both constitutively high expression
        of genes as well as highly variable protein levels, while small values of $\lambda_{\min}$ guarantee constitutively low expression.
        
3. \textit{In the LD regime, the sensitivity of production rate $J$ to
        $\alpha$ is moderated by  $\lambda_0$ and varies across different values of $\alpha$.}  
        Our theory predicts that 
        for $\beta > \beta^*$ (i.e., provided that the termination rate is sufficiently high), 
        the dynamic range of the initiation rate (i.e., the range of $\alpha$ within which the overall protein production rate $J$ varies with $\alpha$) is given by $(0,\alpha^*)$, where 
        the critical initiation rate $\alpha^*$ is defined in \eqref{eq:alphastar}.  
        Furthermore, the degree to which $J$ varies with $\alpha$ 
        is fully determined by the elongation rate $\lambda_0$, as shown in \eqref{eq:J_L}.
        Indeed, $\lambda_0$ controls the time spent by particles at the start of the
        lattice, and can induce significant buffering if $\alpha$ is 
        large enough, thereby modulating the effective rate of entrance
        associated with $J$. We illustrate this in \fref{fig:sensitivity}A,
        where we compare how the current varies as a function of $\alpha$ for
        different values of $\lambda_0$ relative to $\lambda_{\min}$.    
        Recall that the critical initiation rate $\alpha^*$ satisfies the inequalities in \eqref{eq:alphastar_range}, and that $\alpha^*$ increases as $\lambda_0$ decreases.
        \fref{fig:sensitivity}A also shows that for $\lambda_0$ fixed, the production rate of a system closer to the MC regime
        (i.e., with $\alpha$ just below $\alpha^*$)
         is less sensitive to changes in $\alpha$, and that this effect is more pronounced the closer $\lambda_0$ is to $\lambda_{\min}$.
        More generally, the $\alpha$-sensitivity of $J$ increases as $\lambda_0$
        increases.  
        While the dependence of $J$ in $\alpha$ is sublinear
        for $\lambda_0 = \lambda_{\min}$, it becomes linear as $\lambda_0$ gets
        large (see \eqref{eq:J_L}).  This suggests in particular that changes in the free ribosome pool (changing the initiation rate globally) can impact the protein production rate differently across different genes. 

4. \textit{Positioning $\lambda_{\min}$ close to the start site can reduce the amount of ribosomes used.} At maximum production capacity (MC regime), 
        we have shown that the density profile follows the high
        density branch from the start of the lattice until the location $x_{\min}$ of $\lambda_{\min}$ whereafter it adopts the low density branch. This characteristic branch switching phenomenon
        makes $x_{\min}$ critical for the purpose of resource allocation. In \fref{fig:sensitivity}B, we illustrate how a small local change in the rate function can induce a large increase of average particle density when $x_{\min}$ changes substantially. Therefore, a way to limit the excessive usage of ribosomes induced by traffic jams at maximum capacity is to position the minimum rate close to the start. However, as previously shown, positioning it too close to the start (such that $\lambda_0 = \lambda_{\min}$) would also decrease the sensitivity of the system to $\alpha$. 

\subsection*{Empirical Study: Translational Efficiency in Yeast}
\label{sec:yeast}
In light of the aforementioned principles, we explored the extent to which the
translational system in yeast is efficient.  For this study, we used elongation rates 
previously inferred from ribosome profiling data 
for a set of 850 genes in \emph{S. cerevisiae}
\citep{daoduc2018impact} (see \sref{sec:data_processing} of STAR Methods).  These genes were selected in \citet{daoduc2018impact} based on length and footprint coverage, to yield robust estimates of rates. 
The advantage of using this particular dataset over most others
lies in the fact that the inferred rates for this subset of genes
faithfully reproduce ribosome profiling data, incorporating several experimental artifacts of ribo-seq such as undetected stacked ribosomes, thereby minimizing confounding from technical biases.   Furthermore, primarily analyzing high-coverage (and thus likely
highly expressed) genes does not confound our study of design principles, but rather 
provides us an increased signal-to-noise ratio, as these genes are precisely those on 
which our design principles are expected to act most strongly.

We analyzed 
the location of these 850 genes in the phase diagram, and the distribution of
the key parameters and variables that determine the ribosomal currents and
densities.  We found the aforementioned theoretical design principles being reflected as follows:

\begin{figure}[p]
    \centering
    \includegraphics[width=.97\linewidth]{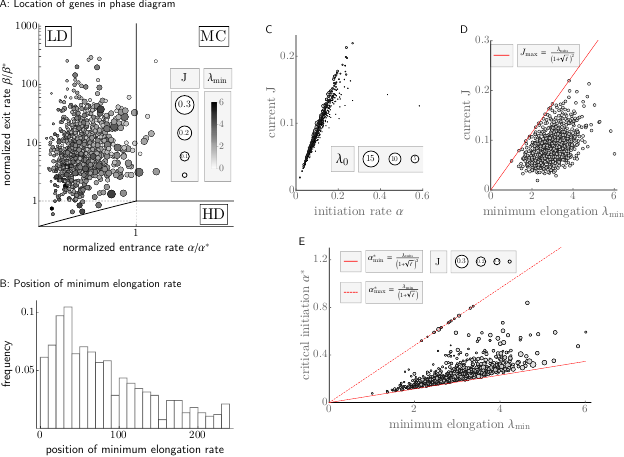}
    \caption{\textbf{Translation machinery in \textit{S. cerevisiae} optimizes
    for ribosomal cost, flexible regulation and production capacity.}
    All rates are in codons per second, while currents are measured in ribosomes per second.
    \textbf{A:} 850 genes of \textit{S. cerevisiae} are located in the phase
    diagram, with size and hue of each data point reflecting current and minimum
    elongation rate, respectively. On a population level, systems of comparable
    production capacities ($\propto\lambda_{\min}$) fully exploit their dynamic
    range by adjustment of $\alpha$, with highly expressed proteins likely
    situated inside or close to MC. \textbf{B:} The resulting resource cost
    considerations drive a significant number of transcripts to position their
    minimum elongation rate early on in the codon sequence, forcing ribosomal
    traffic jams to remain short. 
    \textbf{C:} Initiation $\alpha$ is the main determinant of currents, at least for
    low to average current genes. For highly expressed genes, the correlation between
    $\alpha$ and $J$ decreases due to stronger variation in $\lambda_0$ and
    transitions into MC.
    \textbf{D:} Genes utilize the full dynamical range of currents set by
    $\lambda_{\min}$, through variation in $\alpha$ and $\lambda_0$.
    Constitutively highly expressed genes tend to be closer to maximum
    capacity (red line), while genes with variable expression demands are
    distributed more broadly (see main manuscript). 
    \textbf{E:} For fixed production capacity $\propto\lambda_{\min}$,
    $\alpha^{\ast}$ (the critical initiation rate at which genes reach maximum 
    production capacity) tends to be smaller for genes with larger
    production rates. That is, larger $\lambda_0$ (which are inversely
    related to $\alpha^{\ast}$ for fixed $\lambda_{\min}$) seem to
    facilitate attainment of large currents. Moreover, within highly
    expressed genes, those associated with variable expression patterns over
    time exhibit higher sensitivities (smaller $\alpha^{\ast}$), whereas
    genes with constitutive high expression are found closer towards maximal
    insensitivity (dotted red line) as these configurations ease stable
    expression.
}
    \label{fig:data}
\end{figure}

1. \textit{Translation mainly operates in LD regime.} Upon computing $\alpha^*$
and $\beta^*$, we located the position of each gene in the phase diagram (see
\fref{fig:data}A). Over the 850 genes in our dataset, we found 841 in LD and the
remaining 9  in the MC region. No genes were found in HD, suggesting no
excessive usage of ribosome to achieve any protein level. As a result, the
initiation rate is the main determinant and limiting factor of the current
(Spearman's rank correlation coefficient $\rho = 0.979$). The strength of this
correlation nevertheless decreases as genes get closer to the MC regime, since
$J$ becomes less sensitive to $\alpha$ and $\lambda_{\min}$ becomes its rate
limiting factor (see \fref{fig:data}C). To quantify this reduction in correlation, we binned the data by quartiles of $J$ and computed Spearman correlations within each bin, which yielded (in order of quartiles): $0.93, 0.72, 0.64,$ and $0.58$.

2. \textit{Wide ranges of currents are covered within production capacity.}
For each gene in our dataset, we examined the maximal protein production rate, which according to our theory is proportional to $\lambda_{\min}$. The data 
exhibit an overall range of $\lambda_{\min}$ between $1.01$ and $6.01$ codons/second,
and for any fixed $\lambda_{\min}$, currents are well spread out across
$[0,J_{\max}]$ (see \fref{fig:data}D).
Given that genes cover almost all of the theoretically possible range of currents, we
investigated whether certain configurations of $\lambda_{\min}$ and $J$ are associated with
the biological function of specific genes. 
To do so, we compared 
ribosomal protein genes (known to be highly expressed) and 
genes related to stress response (requiring variable expression over time, see
\sref{sec:data_processing} of STAR Methods).
We found that, while both sets of genes display comparable $\lambda_{\min}$,
ribosomal genes are more likely to be close to their maximal production capacity 
($p < 7\times 10^{-3}$, see \sref{sec:quantification} of STAR Methods) and more consistently so 
(the coefficient of variation is 0.22 for ribosomal genes and 0.36 for stress response).

3. \textit{$\lambda_0$ (associated with sensitivity to $\alpha$) is higher for genes that are either highly expressed or subject to varying expression demand.} 
The impact of increasing $\alpha$-sensitivity is primarily twofold: First, for
fixed production capacity, large
currents may be attained with smaller initiation rates; and second, more substantial changes
in currents may be achieved with small changes in $\alpha$. To investigate the
former we computed $\alpha^{\ast}$, the critical rate necessary for a gene to
attain maximum capacity, across all genes whose $\lambda_{\min}$ exceeded the median
$\lambda_{\min}$ of the data set (as large currents presuppose large capacities).
Further binning this range into quartiles (to isolate the dependence
of $\alpha^{\ast}$ on $\lambda_{0}$), we found that genes whose currents are
at least $90\%$ of the production capacity are significantly more sensitive ($p
< 0.008, 0.01, 0.05,$ and $0.004$, respectively; see
\fref{fig:data}E), requiring smaller initiation rates to reach peak production
rate (cf. \fref{fig:data}C). To inspect the second aspect of
$\lambda_0$ as facilitator or inhibitor of rapid changes in current, we explored the ratio
of $\lambda_0$ to $\lambda_{\min}$ again in ribosomal and stress response genes. For
constitutively highly expressed genes like ribosomal genes, we expect this ratio to be
small to maintain stable current close to MC (cf. \fref{fig:sensitivity}), whereas genes
with variable expression demands like the ones associated with stress response should exhibit
larger ratios. Confirming this intuition, we found significantly reduced levels of 
$\lambda_0 / \lambda_{\min}$ in ribosomal genes ($p < 2\times 10^{-6}$), and significantly
increased levels in stress response genes ($p < 0.04$).

4. \textit{The position of $\lambda_{\min}$ is preferentially located early in the open reading frame.} 
        Upon analyzing the distribution of $x_{\min}$ from our dataset 
        (see \fref{fig:data}B), we found it preferentially located in the codon positions between 30 to 40,
        consistent with genes forestalling excessive ribosome usage through enforcing branch
        switching early on. More specifically, we reasoned that both genes closer to MC and those highly sensitive to $\alpha$ run higher risk of incurring substantial ribosome 
        cost and should thus locate $x_{\min}$ early in the coding sequence.  Indeed,
        both the top quartile of genes close to MC (as measured by $\alpha/\alpha^{\ast}$)
        and stress response associated genes showed significantly smaller $x_{\min}$
        ($p < 0.03$ and $0.01$, respectively). Moreover, genes with unusually large
        values of $x_{\min}$ are significantly less likely to be close to MC (top quartile
        of $x_{\min}$: $p< 1\times 10^{-3}$).

\medskip
To check for systematic biases potentially present in our subsampled gene set and to show replicability of our main biological conclusions, we also analyzed two other independent (and much larger) datasets from 
\citet{williams2014targeting} (combined with polysome profiling from 
\citet{mackay2004gene}) and 
\citet{pop2014} (see Data Processing of STAR Methods). 
We inverted the solution of \eqref{eq:first_order} to obtain approximate estimates of initiation rates, termination rates, and smoothed elongation rates for these datasets, and repeated our analyses. 
As shown in Figure~S4, the results are generally in excellent agreement with what is discussed above (\fref{fig:data}A,B).

\section*{DISCUSSION}
While past quantitative studies of the TASEP under general conditions of extended particle size and/or rate heterogeneity have mostly been limited 
to numerical simulations or mean-field approximations,  
\citep{lakatos2003totally, shaw2003totally, shaw2004mean,chou2004clustered,dong2007inhomogeneous}, 
we used here a different approach that relies on studying  the hydrodynamic limit of the process.
In the case of homogeneous rates, previous studies \citep{schonherr2005, schonherr2004} established this hydrodynamic limit, but without further analyzing the subsequent PDE. After deriving this limit for inhomogeneous rates, we obtained closed-form formulas for the associated current, densities, and phase diagram, 
generalizing previous theoretical results for the TASEP \citep{derrida1993, blythe2007nonequilibrium} and its variants \citep{shaw2003totally, chou2004clustered, stinchcombe2011smoothly}.
Our approach has the advantage of revealing the key parameters that the current and densities depend on, enabling an immediate quantification of the process and its phase diagram. Such a quantification is difficult to
achieve via conventional stochastic simulations or approximations used in the past several years \citep{zia2011modeling, zur2016predictive, szavits2018deciphering}.

Our characterization of the current and densities in the phase diagram  suggests that, in 
agreement with earlier experimental studies  \citep{kosuri2013composability, 
salis2009automated},  translation dynamics should be mainly governed by the initiation rate,
while the termination rate and most elongation rates have negligible impact. In 
particular, our results explain why having the initiation rate as the main limiting factor 
of the current \citep{plotkin2011synonymous} minimizes ribosome usage. 
In addition, we discovered the importance of smoothed rather than raw elongation
profiles in predicting translation dynamics, explaining the previously observed
mild effect that any individual elongation change has compared to accumulated,
neighboring changes \citep{levin2018genome}.  This allowed us to identify two key parameters 
of the system, namely, the smoothed elongation rate $\lambda_0$ immediately following initiation and the
minimal smoothed elongation rate $\lambda_{\min}$.  Previous studies have established some association 
between the sequence context in the early $5'$ coding region and protein production 
levels \citep{frumkin2017gene, boel2016, ben2015rationally}. For example, it has been shown 
that mRNA secondary structure in the first $\sim 16$ codons (which locally decreases the 
elongation rate) negatively affects the translation rate  in \emph{E. Coli}, while no 
significant contribution of mRNA folding in other regions was found \citep{frumkin2017gene}. 
By exposing $\alpha$ and $\lambda_0$ as the only parameters that currents in LD depend on, 
our analysis suggests a direct explanation for such contrast.

We also highlighted the impact of $\lambda_0$ on the sensitivity of the current to changes in $\alpha$. In practice, initiation rates can vary at the individual gene level (e.g., through interactions with specific miRNAs \citep{humphreys2005micrornas}). According to our theory, the way that these variations impact the protein production rate depends on $\lambda_0$; we hence suggest that this may explain why genes associated with stress response present higher values of $\lambda_0$, as it facilitates the response to changes in $\alpha$. 
At a more global level, our study shows how protein levels can be more or less
robust against changes in the ribosomal pool, which can simultaneously affect
all initiation rates in a cell \citep{shah2013}. Since the level of ribosomes
present in a cell fluctuates over time \citep{wyant2018nufip1}, it would be
interesting to see if protein levels scale uniformly with these variations
across genes, and if not, whether the differences in $\lambda_0$ can explain it.

To the best of our knowledge, the role of the minimum elongation rate $\lambda_{\min}$ has so far received 
attention only indirectly, through the study of what is known as the ``$5'$
translational ramp'' \citep{tuller2010an}. This ramp is a pattern of
translational slowdown around codon position $30$-$50$ followed by steadily
accelerating elongation rates, which is mirrored by the spatial distribution of
minimum elongation rates we found here.
This ramp has been hypothesized to prevent
crowding of ribosomes on the transcript \citep{tuller2010an}, for which we provide a theoretical basis, exposing $\lambda_{\min}$ as a separator between crowded and freely elongating ribosomes. 
More generally, the complex interplay between the maximum current capacity, ribosome usage,
and sensitivity to the initiation rate suggests various ways to set the parameters $\lambda_0$, $\lambda_{\min}$ and $x_{\min}$, depending on the desired object to optimize. For example, allocating the minimum elongation rate near the beginning of the ramp region provides an optimal trade-off between high sensitivity and minimal traffic jams. On the other hand, it would be optimal for genes with housekeeping function to 
have a decreased sensitivity, which would push the minimum to earlier positions.

Our analysis can also help to answer the long-debated question regarding the
implication of translation on codon usage bias \citep{hershberg2008selection,
frumkin2018codon, shah2013}. Since highly expressed genes are enriched for synonymous codons
translated by more abundant tRNAs \citep{yu2015codon, hanson2018codon}, it has been 
hypothesized that codon usage bias increases the overall protein synthesis rate by accelerating 
elongation \citep{hershberg2008selection}.  However, recent studies have challenged such a 
hypothesis, suggesting that translational selection for speed is not sufficient to explain 
the observed variation in codon usage bias \citep{mahajan2018translational}. Synonymous 
changes of the coding sequence modify local elongation rates, but, according to our theory, 
such a modification impact the overall protein production rate only if the
smoothed elongation rates $\lambda_0$ or $\lambda_{\min}$ are affected.
In addition, our work implies that synonymous codon replacements that substantially change the location $x_{\min}$ of $\lambda_{\min}$ affect the efficiency of ribosome usage, and hence are more likely to be under selective pressure.
Aside from these cases, there should be little \emph{direct} impact of synonymous codon usage on translation efficiency;  this prediction is consistent with previous studies that tried to explain
differences in expression using codon identity \citep{gustafsson2012engineering},
and to characterize the sensitivity of translational output
with respect to changes in elongation \citep{levin2018genome}.
Codon usage bias could affect the protein production rate \emph{indirectly}, however, by reducing the cost of translation:  
replacing a codon by a ``faster'' synonymous codon
helps to reduce the local ribosome density on the transcript, and this can in turn increase the availability of free ribosomes and therefore increase the initiation rate $\alpha$ slightly; in the LD regime, increasing $\alpha$ would increase the protein production rate.
We note that other factors such as mRNA decay \citep{hanson2018codon}, or reduction of
nonsense errors or co-translational misfolding \citep{gilchrist2007combining,
frumkin2018codon} might be more important drivers of codon usage bias.

Finally, it would be interesting to experimentally test our theoretical predictions, e.g., using cell-free expression protocols such as lysate-based systems, which have been developed to optimize protein synthesis and more recently refined to study translation dynamics \citep{moore2017cell, rosenblum2014engine, katrandis2019}. By designing an appropriate mRNA sequence and controlling different components (NTPs, ribosomes, tRNAs, specific amino acids), these systems allow to manipulate the initiation and elongation rates, and hence tune the key parameters identified by our theoretical analysis. For example, one can modify $\lambda_{\min}$ or $\lambda_0$ by changing the level of corresponding amino acids, and vary $\alpha$ by modifying the $5'$ UTR sequence or changing the ribosome concentration.  The flexible nature of such cell-free expression systems, coupled with precise measurement of protein levels (e.g., via isotope-labeled amino acids or reporter proteins), should help to verify our theoretical results.  In particular, it would be interesting to experimentally demonstrate the existence of phase transitions, and by modifying the mRNA sequence, test our predictions on how to effectively control the robustness and sensitivity of the translation system.  We are currently pursuing these research directions.

\section*{ACKNOWLEDGEMENTS}
This research is supported in part by NIH grants R01-GM094402 and R35-GM134922; a Packard
Fellowship for Science and Engineering; and the Koret--UC Berkeley--Tel Aviv University Initiative in Computational Biology and Bioinformatics. 
YSS is a Chan Zuckerberg Biohub Investigator. 


\section*{DECLARATION OF INTERESTS}
The authors declare no competing interests.

\clearpage

\section*{STAR METHODS}\label{sec:appendix}
\subsection*{LEAD CONTACT AND MATERIALS AVAILABILITY}
Further information and requests for resources and reagents should be directed to and 
will be fulfilled by the Lead Contact, Yun S. Song (yss@berkeley.edu).
\\[2mm]
This study did not generate new reagents.

\subsection*{METHOD DETAILS}
\subsubsection*{The hydrodynamic limit of the inhomogeneous $\ell$-TASEP}
\label{sec:methods_hydro_limit}
We derive here the PDE governing the hydrodynamic limit of the open-boundaries inhomogeneous 
$\ell$-TASEP. To do so we exploit a representation of its dynamics in terms of another 
interacting particle system, the so-called zero range process (ZRP), whose hydrodynamics can
be found explicitly. This TASEP-ZRP duality provides an expedient and general
tool for identifying explicit TASEP formulas; however, rigorously proving the
validity of these formulas often requires more technical tools from probability
theory. Since this work's emphasis is on the application of TASEP to
unraveling the key parameters of translation dynamics, we will here concentrate on
showcasing the TASEP-ZRP framework, and keep a rigorous existence proof of the
hydrodynamic limit, combining techniques from
\citet{rezakhanlou1991hydrodynamic,covert1997hydrodynamic} and
\citet{bahadoran2012hydrodynamics}, to a separate manuscript.

\paragraph{\it Reduction to periodic boundaries and mapping to the ZRP.}
The purpose of the hydrodynamic limit is to describe the local evolution of the
macroscopic particle density in the large system limit.
As such, it does not explicitly rely on the precise formalism by which particles 
enter and exit the lattice at the boundaries (which will only later be needed to
impose boundary conditions on the resulting PDE
\citep{bahadoran2012hydrodynamics}). In particular, we are free to
choose periodic boundary conditions for our limiting procedure without changing the
resulting PDE \citep{schonherr2004}. This has the advantage of preserving the
total number of particles, which is essential for establishing the
correspondence between TASEP and ZRP. In
the following, we thus consider the $\ell$-TASEP
with $M$ particles on a \textit{ring} of $N$ sites jumping to the right at rate $p_i$, 
and take $M,N\to\infty$ while $M/N$ remains constant.

The ZRP is now obtained by reversing the roles of holes and particles: 
It consists of $N-M\ell$ particles (corresponding to the
$N-M\ell$ holes in the TASEP) distributed across $M$ sites (matching the TASEP
particles) $\{1, \ldots, M\}$, with multiple particles allowed to stack up on the 
same site. A ZRP configuration $(\xi_{i,t})_{1\leq i \leq M}$ describes the number 
of particles $\xi_{i,t}$ at each site $i\in\{1,\ldots,M\}$ and time $t$, 
and can be seen as a representation of spacings between particles $i$ and
$i+1$ in the TASEP.

As a result, the TASEP dynamics are translated into ZRP dynamics as follows:
If a site $i$ at time $t$ is occupied by at least one particle, 
then the topmost particle jumps to the left with rate $m_{i,t} = p_{k(i,t)}$, where
$k(i,t)$ is the position of the $i$th TASEP particle (see formula
\eqref{eq:ZRP_to_TASEP} below) at time $t$. This 
jump occurs regardless of whether the destination site is occupied or not. That
is, neither exclusion nor long range interactions are present, which will be key to 
establishing the hydrodynamic limit.

The correspondence between TASEP and ZRP states described above is so far only
determined up to rotations of the TASEP lattice, hence we introduce one further 
variable $\xi_{0,t}\in\{1,\ldots,N\}$ to trace the position of particle $1$. 
More explicitly, at time $t$, TASEP particle $i$ is located at site
\begin{equation}
    k(i,t) = \sum_{j=0}^{i-1} \xi_{j,t} + \ell(i-1)
    \label{eq:ZRP_to_TASEP}
\end{equation}
on the TASEP ring. An illustration of this correspondence is given in
Figure~S1.

\paragraph{\it The hydrodynamic limits of the ZRP and TASEP.}
The connection between the TASEP and the ZRP has been fruitfully used to derive
hydrodynamic limits for homogeneous systems \citep{schonherr2004,
schonherr2005}. Here we generalize this approach to heterogeneous lattices and
supply appropriate boundary conditions to the PDE, which become necessary when
working with open rather than periodic boundaries.

We start with the master equation associated with the ZRP:
\begin{equation}
    \partial_t \xi_{i,t} = m_{i+1,t}z_{i+1,t} - m_{i,t}z_{i,t},
    \label{eq:ZRP_master}
\end{equation}
where $z_{i,t} = \mathbb P(\xi_{i,t} > 0)$ is the probability that site $i$
is non-empty at time $t$. Our goal is to identify a PDE that describes the limit
of \eqref{eq:ZRP_master} under Euler scaling, i.e., on time scale $at$ and
spatial scale $ia$. Denoting these scaled variables as $t$ again in time and $x,
y$ in space such that $k = \lfloor x/a \rfloor$ and $i = \lfloor y/a \rfloor$, and assuming the existence of a continuously differentiable rate
function $\lambda$ such that $\lambda(x) = p_k$, the master equation
\eqref{eq:ZRP_master} becomes
\begin{align}
    a\partial_t c(y,t) &= \lambda(x(y+a,t)) z(y+a,t) - \lambda(x(y,t)) z(y,t)
    \notag    \\
    &= a \partial_y [\lambda(x(y,t)) z(y,t)] + \dfrac{a^2}{2} \partial_{yy} [
    \lambda(x(y,t)) z(y,t) ] + O(a^3),
    \label{eq:ZRP_pre_hydro_limit}
\end{align}
where $c(y,t)$ and $z(y,t)$ are the continuum limits of $\xi_{i,t}$ and
$z_{i,t}$, respectively. Under local stationarity \citep{kipnis2013scaling},
we may replace $z$ in \eqref{eq:ZRP_pre_hydro_limit} using the fugacity-density 
relation $z = c(1+c)^{-1}$ to obtain the final hydrodynamic limit of the
inhomogeneous ZRP as
\begin{equation}
    \partial_t c = \partial_y\bigg( \lambda \dfrac{c}{1+c} \bigg) + \dfrac{a}{2}
    \partial_{yy} \bigg( \lambda \dfrac{c}{1+c} \bigg).
    \label{eq:ZRP_hydro_limit}
\end{equation}
The assumption of local stationarity is essentially justified by the
one-block estimates in \citet{covert1997hydrodynamic}, as long as one can ensure
slow enough variation of $\lambda(x(y,t))$ in $t$. In our case, this smooth
dependency is given, since in a small (on the Eulerian scale) time interval
$N\Delta t$, we expect a particle to perform $O(N\Delta t)$ jumps, and whence
$\lambda(x(y,t+N\Delta t)) - \lambda(x(y,t)) \in O(\Delta t)$.

To derive the corresponding PDE for the TASEP, we use \eqref{eq:ZRP_to_TASEP} to establish the
continuum relation between $x,y$ and $t$. More precisely,
\begin{equation}
    x(y,t) = ak(i,t) = a\bigg(\sum_{j=0}^{i-1} \xi_{j,t} + \ell(i-1)\bigg)
    = \int_0^y c(u,t) \ \mathrm{d}u - \dfrac{a}{2}\bigg( c(y,t) - c(0,t) \bigg)
    + \ell(y-a) + O(a^2).
    \label{eq:continuous_ZRP_to_TASEP}
\end{equation}
Upon recognizing that particle densities 
are related by $\rho = (c+\ell)^{-1}$ and changing coordinates according to 
\eqref{eq:continuous_ZRP_to_TASEP},  \eqref{eq:ZRP_hydro_limit} yields the hydrodynamic limit of the TASEP
\begin{equation}
\partial_t \rho = -\partial_x\left[\lambda(x) \rho G(\rho) \right] -
            \frac{a}{2} \partial_{xx} \left[ \lambda(x)G(\rho) \right] + O(a^2),
            \label{eq:app_hydro_limit}
\end{equation}
where $G(\rho) = \frac{1-\ell\rho}{1-(\ell-1)\rho}$.
\subsubsection*{Phase diagram analysis}
\label{sec:phase_diagram_analysis}
We now use \eqref{eq:app_hydro_limit} to provide a detailed derivation of the phase
diagram described in the main text.
\paragraph{\it Reduction to conservation law.}
Solutions of \eqref{eq:app_hydro_limit} converge locally uniformly (under mild
conditions on $\lambda$, see \sref{sec:phase_transition}) to viscosity solutions of
the scalar conservation law
\begin{align}
    \partial_t \rho(x,t) &= -\partial_x[
    \underbrace{\lambda(x)H(\rho(x,t))}_{J(\rho(x,t),x)} ],
    \label{eq:app_first_order}
\end{align}
where $H(\rho) = \rho G(\rho)$, which thus determines the phase diagram in the 
hydrodynamic regime. Setting $\partial_t\rho = 0$ identifies the stationary profiles 
of the TASEP as distributions satisfying
\begin{equation}
    J\left(\rho,x\right) = J_c,
    \label{eq:stationary_first_ode}
\end{equation}
where $J_c = J_c(\alpha,\beta,\lambda)$ is the critical current, set to belong to $[0,J_{\max}]$, where $J_{\max}$ is the transport capacity of the lattice
\begin{equation}
    J_{\max} = \min_{x\in[0,1]}\max_{\rho\in[0,1/\ell]} J(\rho,x) =
    \dfrac{\lambda_{\mathrm{min}}}{(1+\sqrt{\ell})^2}.
    \label{eq:transport_capacity}
\end{equation}
 \eqref{eq:stationary_first_ode} has two solutions (see
Figure~S5A) of the form
\begin{equation}
    \rho_{\pm}(x) = \dfrac{1}{2\ell} + \dfrac{J_c(\ell-1)}{2\ell\lambda(x)} \pm
    \sqrt{ \bigg( \dfrac{1}{2\ell} + \dfrac{J_c(\ell-1)}{2\ell\lambda(x)} \bigg)^2 -
    \dfrac{J_c}{\ell\lambda(x)}},
    \label{eq:stationary_solutinos}
\end{equation}
any mixture of which may be a potential attractor picked by the system as
$t\to\infty$. Deciding precisely which mixture dominates requires analysis of 
the characteristic curves. 

\paragraph{\it Solving the characteristic ODE.}
\label{sec:characteristic}
Denoting the characteristic curves by $x^t$ and $\rho^t$ with initial data
$x^0, \rho^0$, their evolution is described by the system of ODE
\citep{evans2010partial}
\begin{align}
        \frac{dx^t}{dt} &= \lambda(x^t) H'(\rho^t),
        \label{eq:app_incharacteristics1} \\
        \frac{d\rho^t}{dt} &= - \lambda'(x^t) H(\rho^t),
        \label{eq:app_incharacteristics2}
\end{align}
where $H'$ and $\lambda'$ respectively denote the derivatives of $H$ and $\lambda$ with respect to their arguments.
The solutions are easily verified to be
\begin{align}
    x^t &= F^{-1}(t) \label{eq:incharacteristics1sol} \\
    \rho^t &= H^{-1}\left( \dfrac{J(\rho^0,x^0)}{\lambda(x^t)} \right)
    \label{eq:incharacteristics2sol}
\end{align}
as long as $J(\rho^0, x^0) \in [0,J_{\max}]$. The form of $F$ follows from
formally separating variables:
\begin{equation}
    F(x) = \int_{x^0}^x \dfrac{1}{ \lambda(y) H' \circ H^{-1}(J(\rho^0,x^0)/
    \lambda(y)) } \ \mathrm dy,
    \label{eq:defineF}
\end{equation}
while $H^{-1}(J(\rho^0,x^0)/\lambda(x^t))$ is understood 
to be the preimage compatible with $\rho^0$, see Figure~S5A. For
the homogeneous $\ell$-TASEP \eqref{eq:incharacteristics1sol} and
\eqref{eq:incharacteristics2sol} depend linearly on each other, giving rise to
straight line characteristic curves (see Figure~S5B). In
the more general heterogeneous setting, however, more complicated behavior
emerges (Figure~S5C). In particular, if $J(\rho^0,x^0) < J_{\max}$,
then for all $t\geq 0$, 
\[
    \frac{J(\rho^0,x^0)}{\lambda(x^t)}  < \frac{1}{(1+\sqrt{\ell})^2},
\]
so $\rho^t < \frac{1}{\ell + \sqrt{\ell}}$ for all $t$ if $\rho^0 < \frac{1}{\ell + \sqrt{\ell}}$, while
$\rho^t > \frac{1}{\ell + \sqrt{\ell}}$ for all $t$ if $\rho^0 > \frac{1}{\ell + \sqrt{\ell}}$.
Hence, the sign of $\frac{dx^t}{dt} =\lambda(x^t) H'(\rho^t)$ remains the same for all $t$, 
and any characteristic
curve $x^t$ starting at the left lattice boundary $x^0 = 0$ or right lattice boundary
$x^0 = 1$ propagates towards the opposite end and fills the lattice entirely. 

On the other hand, if $J(\rho^0, x^0) > J_{\max}$, then $\frac{J(\rho^0,x^0)}{\lambda(x_{\min})} > \frac{1}{(1+\sqrt{\ell})^2}$,
where $x_{\mathrm{min}} = \arg\min_x \lambda(x)$, so $H^{-1} \Big(\frac{J(\rho^0,x^0)}{\lambda(x_{\min})}\Big) > \frac{1}{\ell}$.
Recalling \eqref{eq:incharacteristics2sol} and noting that it is physically not possible to have $\rho^t > \frac{1}{\ell}$, we conclude that 
the characteristic curve $x^t$ cannot reach $x_{\min}$.  
Indeed, it follows from \eqref{eq:app_incharacteristics1} and 
\eqref{eq:app_incharacteristics2} 
that at some critical time $t_c$ before reaching $x_{\min}$,
 the characteristic curve $x^t$ reverses direction 
 while $\rho^t$ crosses
$\arg \max_{\rho} H(\rho) = (\ell+\sqrt{\ell})^{-1}$, resulting in $x^t$ 
returning to its origin. 
 \fref{fig:illustration}E
of the main text and
Figure~S5D illustrate this behavior.

\paragraph{\it Computing initial densities $\rho^0$.}
As a consequence of the above, determining phase transitions in the $\alpha$-$\beta$ 
phase diagram reduces to establishing regimes in which $J(\rho^0,x^0)$ exceeds or falls 
short of $J_{\max}$, which in turn is equivalent to finding an expression for $\rho^0$ in
terms of $\alpha$ and $\beta$. This is done by considering each lattice
end separately and balancing currents:

\textit{The right lattice end $x^0=1$:}
As described in the main text, $\rho_1 = \rho(1)$ decomposes into a sum of two 
contributions, the periodic part $\rho_1^+$ and the troughs $\rho_1^-$ 
\citep{chou2004clustered}. More explicitly,
\begin{equation}
    \rho_1 = \dfrac{1}{\ell}\bigg[ (\ell-1)\rho_1^- + \rho_1^+ \bigg].
    \label{eq:split_density}
\end{equation}
Since the current $J_c$ is a conserved quantity of the system, the local
currents across the last lattice site, the second to last lattice site and
within the last $\ell$ sites must all be the same:
\begin{equation}
    J_R := J(\rho_1, 1) = \beta\rho_1^+ = \lambda_1 \rho_1^-.
    \label{eq:balance_right_current}
\end{equation}
Solving for $\rho_1$ gives exactly $\frac{1}{\ell}(1 - \frac{\beta}{\lambda_1})$. 
Consequently, $J_R \leq J_{\max}$ iff
\begin{equation}
    \beta < \beta^{\ast} = \frac{1}{2} \Bigg[ \lambda_1 -
            \frac{\ell-1}{(1+\sqrt{\ell})^2} \lambda_{\mathrm{min}}   
            \label{eq:betastar}    -\sqrt{ \left(
                    \lambda_1 - \frac{\ell-1}{(1+\sqrt{\ell})^2}
                \lambda_{\mathrm{min}} \right)^2 - \frac{4 \lambda_1
        \lambda_{\mathrm{min}} }{(1+\sqrt{\ell})^2} }\; \Bigg].     
   \end{equation}

\textit{The left lattice end $x^0=0$:}
Computing $\alpha^{\ast}$ is more delicate as the effective jump rate is a
combination of entrance rate and particle exclusion. To bypass this problem, we
investigate the current of \textit{holes} rather than particles, which is running in
the opposite direction.
With the loss of the particle-hole symmetry present in the simple $1$-TASEP
\citep{derrida1993}, the hole density $\rho^h$ here assumes a more complicated form. It
satisfies its own conservation law given by
\begin{equation}
    \partial_{t^h} \rho^h = \partial_x [J^h(\rho^h, x)],
    \label{eq:hole_PDE}
\end{equation}
where
\begin{equation}
    J^h(\rho^h, x) = \lambda(x)\rho^h \dfrac{1-\rho^h}{1+(\ell-1)\rho^h}
    \label{eq:hole_current}
\end{equation}
and $t^h = \ell t$ is the time scale of the holes, moving slower as their
density is higher. Thus by balancing hole currents rather than particle currents 
at $x^0 = 0$, we obtain, noting that the effective exit rate (of holes) is still
$\alpha$ (as $\ell$ holes need to accumulate for exiting to happen),
\begin{equation}
    J^h(\rho_0^h,0) = \alpha\rho_0^h.
    \label{eq:hole_current_balance}
\end{equation}
Solving for $\rho_0^h$ and using $\rho_0^h  = 1 -\ell\rho_0$, we obtain
 $\rho_0 = \alpha/[\lambda_0+(\ell - 1)\alpha]$. 
Defining $J_L := J(\rho_0,0)$, we obtain $\alpha^{\ast}$ by solving for  $\alpha, \ J_L = J_{\max}$.

\paragraph{\it Phase transitions and profiles.}
\label{sec:phase_transition}
Using the densities obtained from \eqref{eq:balance_right_current} and
\eqref{eq:hole_current_balance} in the characteristic curves
\eqref{eq:app_incharacteristics1} and \eqref{eq:app_incharacteristics2} yields the HD
and LD regimes for parameter configurations $(\alpha > \alpha^{\ast}, \beta <
\beta^{\ast})$ and $(\alpha < \alpha^{\ast}, \beta > \beta^{\ast})$, respectively. 
To describe the phase transition between HD and LD, we observe that for $\alpha <
\alpha^{\ast}$ and $\beta < \beta^{\ast}$ both characteristic curves move into
the lattice, meet, and move along a common shock with speed 
\begin{equation}
    v_{\mathrm{shock}} = \dfrac{J_R - J_L}{\rho_r - \rho_l},
    \label{eq:shock_speed}
\end{equation}
where $\rho_l$ and $\rho_r$ are the densities left and right of the shock.
As $\rho_{r} - \rho_{l} > 0$ as long as $\alpha < \alpha^{\ast}$ and 
$\beta < \beta^{\ast}$ (cf. Figure~S5A), $v_{\mathrm{shock}} > 0$ if and only if
$J_R > J_L$. That is, the slower current pushes the faster one
past the lattice boundaries and dominates the stationary behavior of the system.
The HD and LD regimes are thus separated by incoming currents of
equal magnitudes
\begin{equation}
    J_L = \dfrac{\alpha(\lambda_0 - \alpha)}{\lambda_0 + (\ell - 1)\alpha} =
    \dfrac{\beta(\lambda_1-\beta)}{\lambda_1 + (\ell - 1)\beta} =  J_R.
    \label{eq:HD_LD_transition}
\end{equation}

Lastly, we can use the behavior of characteristic curves for $J(\rho^0,x^0) > J_{\max}$ to
describe stationary profiles in the MC regime ($\alpha > \alpha^{\ast}$ and $\beta >
\beta^{\ast}$):  Each characteristic curve reverses direction at a critical time
$t_c$ and returns to its respective lattice boundary, while the density $\rho^t$ it 
carries transitions from  $\rho_-$ to $\rho_+$ (on the left characteristic) or $\rho_+$ to 
$\rho_-$ (on the right characteristic). Since the reversal of directions occurs
strictly before reaching $x_{\min}$, these characteristics
provide density information on only part of the lattice. The 
uncovered regions are determined by the simultaneously propagating rarefaction waves
\citep{evans2010partial}, which
interpolate between $x^t$ and the characteristic curve $x_{\max}^t$ associated
with $J(\rho^0,x^0) = J_{\max}$ (see Figure~S5D). Together,
these observations combine to produce the high density and low density profiles to the left
and right of $x_{\min}$, respectively, with critical current $J_c = J_{\max}$, 
as described in the main manuscript.

If $\lambda$ has exactly one global minimum $x_{\min}$, this description
captures the density profile on the entire lattice. In the case of multiple
global minima at $\{x_{\min,1}, \ldots, x_{\min,n}\}$ however, it describes
$\rho$ on $[0,x_{\min,1}] \, \cup \, [x_{\min,n},1]$ only, leaving open
fluctuations on the middle segment $(x_{\min,1}, x_{\min,n})$. Although unlikely
to be encountered in practice, these singular rate functions exhibit interesting
stochastic phenomena: The presence of
high densities on the initial interval and low densities on the terminal one
suggest the formation of a coexistence phase in-between. Indeed, the subsystem
restricted to $[x_{\min,1},x_{\min,n}]$ may be regarded as a TASEP with entrance
and exit rates $\alpha = \beta = \lambda_{\min}/(1 + \sqrt{\ell})$, positioning
it at the triple point of the phase diagram, and computing the
characteristics reveals one or multiple stationary shock fronts in the interior. Such
macroscopic phenomenon in the homogeneous $1$-TASEP has previously been associated 
on the microscopic level with a shock performing a random walk on the lattice with 
reflecting boundaries \citep{derrida1997shock}. Numerical simulations
seem to locate these shock around local maxima disproportionately often (cf.
Figure~S2), which might reflect dependencies of its diffusivity 
on $\lambda$. 

\subsubsection*{Applicability to discrete lattices}
\label{sec:applicability}
The existence of a continuous limiting rate function $\lambda: [0,1]\to\mathbb
R^+$ extending the discrete jump rates $p_k = \lambda(ak)$ is an important
ingredient in our treatment of the hydrodynamic limit. That is, in order for
density profiles to be accurately approximated by solutions to the PDE
\eqref{eq:first_order}, the $p_k$ must vary smoothly across lattice sites.
Microscopic systems like the translation machinery in cells, however, are
typically subjected to substantial amounts of fluctuations, resulting in far
rougher elongation profiles (see \fref{fig:smoothing}A). Despite this lack of
regularity, the hydrodynamic limit can still be employed to describe
local averages of such a system. More precisely, fixing $r\in \{1, \dots, N\}$, we
associate with an elongation rate profile $\{p_1, \dots, p_N\}$ and the corresponding 
density profile 
$\{\rho_1, \dots, \rho_{N}\}$ their smoothed profiles $\{\overline{p}_1,\dots, 
\overline{p}_{N-r+1}\}$ and $\{\overline{\rho}_1, \dots, \overline{\rho}_{N-r+1}\}$, respectively,
obtained through a moving $r$-codon average: $\overline{p}_k = \sum_{i=k}^{k+r-1} p_i/r$, 
and $\overline{\rho}_k = \sum_{i=k}^{k+r-1} \rho_i/r$. Moreover, we define
$\{\sigma_1, \dots, \sigma_{N-r+1}\}$ to be the steady state density profile
under the elongation rates $\{\overline{p}_k\}$. If $\{p_k\}$ extends to a
smooth $\lambda: [0,1]\to\mathbb R^+$, then since $|\overline{p}_k - p_k| \in
O(N^{-1})$, $\{\overline{p}_k\}$ extends to this same $\lambda$, and hence
$\{\rho_k\}, \{\overline{\rho}_k\}$ and $\{\sigma_k\}$ all converge to the solution
$\rho$ of \eqref{eq:first_order}. When $\{p_k\}$ does not extend to a
continuous limit, then $\{\rho_k\}$ generally does not either. However, by the same 
reasoning that establishes the hydrodynamics for the $1$-TASEP with quenched disorder 
\citep{seppalainen1999existence}, $\{\overline{\rho}_k\}$ should still be close
to $\{\sigma_k\}$, which, due to the greater regularity of
$\{\overline{p}_k\}$, is well approximated by the hydrodynamic density profile 
under $\{\overline{p}_k\}$. Thus, $\{\overline{\rho}_k\}$ is ultimately well 
approximated by the hydrodynamic limit under $\{\overline{p}_k\}$.

To confirm this, we carried out an extensive simulation study
on elongation rate profiles obtained from ribosome profiling data of yeast (see
\sref{sec:data_processing} for more details on data).  Specifically, we
performed the smoothing $\{p_k\} \to \{\overline{p}_k\}$ (\fref{fig:smoothing}A,B), 
simulated density profiles $\{\rho_k\}$ under $\{p_k\}$ (\fref{fig:smoothing}A,C), 
and compared the corresponding smoothed densities $\{\overline{\rho}_k\}$ with
the hydrodynamic prediction under $\{\overline{p}_k\}$ (\fref{fig:smoothing}D). 
A choice of $r = 10$, which is equal to the particle (ribosome) size $\ell$ in 
translation and the smallest window size guaranteeing smoothness of
$\{\overline{p}_k\}$ due to the $\ell$-periodicity induced by traffic jams,
resulted in excellent agreement both in densities and currents uniformly across 
transcripts while maintaining local structure.  

\subsubsection*{Boundary conditions}
\label{sec:boundary_conditions}
The computation of initial densities in \sref{sec:characteristic} yielded precise 
boundary values for $x = 0$ in the LD regime and $x = 1$ in the HD regime, respectively.
Using the same principle of balancing currents, boundary conditions for all
locations in the phase diagram can be computed. The results are listed in
Table~S1, which extend previous results obtained in
\citep{lakatos2003totally} (who derived entries (1,1), (2,2) and (2,3) of Table~S1).
More precise information about the boundary layers can be gleaned
from direct analysis of \eqref{eq:app_hydro_limit} rather than its limit
\eqref{eq:app_first_order}. 

\subsubsection*{Data processing}
\label{sec:data_processing}
Initiation, elongation, and termination rates were obtained from an earlier work
\citep{daoduc2018impact}, where the rates were estimated from ribosome profiling
data of \emph{S. cerevisiae} for a set of 850 genes selected based on length and
footprint coverage. 
The initiation and termination rates ($\alpha$ and $\beta$) were taken directly from that previous work.  
To compute the 
elongation rates relevant to the hydrodynamic limit, we applied a ten-codon moving average 
to their elongation rates (see \sref{sec:applicability}). 
To demonstrate replicability on larger datasets, we took ribosome 
profiles directly from \citet{williams2014targeting} and \citet{pop2014}
(combined with polysome profiling from \citet{mackay2004gene} for normalization
purposes, yielding 3098 and 2536 genes, respectively), smoothed them by moving averages of length $\ell = 10$, and inverted the
solution of \eqref{eq:first_order} to obtain initiation rates, termination
rates, and smoothed elongation profiles.

\subsection*{QUANTIFICATION AND STATISTICAL ANALYSIS}
\label{sec:quantification}
\subsubsection*{Hypothesis tests and $p$-values}
\label{sec:p-values}
To establish significance of a subset $X$ of genes with respect to a statistic
$f$ (e.g., $\alpha, J$ or $x_{\min}$) relative to a background set $Y$, we performed 
hypothesis testing on the median $m_f$ of $f$ over samples in $X$.
Under the null distribution of $X$ being drawn uniformly at random, the
probability of this test statistic exceeding $m$ equals the probability of a 
hypergeometric variable with parameters $N=\left| Y \right|, K= 2\left| Y_m \right|, 
n=\left| X \right|$, where $Y_m$ is the set of genes in $Y$ whose $f$ exceeds
$m$, exceeding $\lfloor \left| X/2 \right| \rfloor$. This p-value can be computed
explicitly. Sets of ribosomal and stress response genes were taken from the Saccharomyces
Genome Database \citep{cherry2011saccharomyces}.

\subsubsection*{Agreement between theoretical prediction and simulation}
\label{sec:correlations}
In order to empirically verify our theoretical justification of the hydrodynamic
limit, we simulated ribosome profiles and currents for all 850  \emph{S. cerevisiae} genes studied in \citet{daoduc2018impact}.  For each gene, we considered four conditions: LD,
HD, MC, and under the actual initiation and termination rates inferred in
\citet{daoduc2018impact}; these four conditions correspond to different rows in
Figure~S6.  Absolute errors in ribosome density profiles and currents (first and
last columns of Figure~S6) are accurately predicted across all gene lengths---with a slight
increase in prediction accuracy for longer genes (as expected, since the
hydrodynamic limit becomes exact in the infinite length limit)---and across all regimes 
of the phase diagram. Due to two or more bottlenecks occasionally competing on the same transcript
(i.e., when $|\{x \ : \ \lambda(x) = \lambda_{\min}\}| > 1$, cf. last paragraph of
\sref{sec:phase_transition} of STAR Methods), error distributions in MC exhibit heavier 
tails than in LD and HD.  However, overall these outliers do not affect the 
quality of our theoretical prediction significantly. In particular, correlations
between simulated and theoretical transcript-by-transcript quantities---ribosome density profiles and mean occupancies (middle column), as well as currents (last column)---are consistently
high, demonstrating good predictive power of our hydrodynamic framework.  

In HD, predicted and simulated ribosome density profiles had quite low mean
squared differences (second row, first column of Figure~S6), but poor correlation (histograms in second row, second column).  
This seemingly contradictory result can be explained by 
typical fluctuations in theoretical density profiles being of
the same order as typical fluctuations in the random noise (mean ratio of fluctuations 
$=0.037$). That is, generic HD profiles are close to flat, allowing uncorrelated
site-by-site noise to substantially reduce overall correlations.

\subsection*{DATA AND CODE AVAILABILITY}

This study did not generate new data.  Code, including the code used to generate all 
figures, is publicly available at \url{https://github.com/songlab-cal/l-TASEP}.

\clearpage
\bibliographystyle{apalike}
\bibliography{bibliography}{}


\clearpage

\section*{Supplementary Figures and Table}
\beginSupplement

\begin{figure}[h]
    \centering
    \includegraphics[width=.5\textwidth]{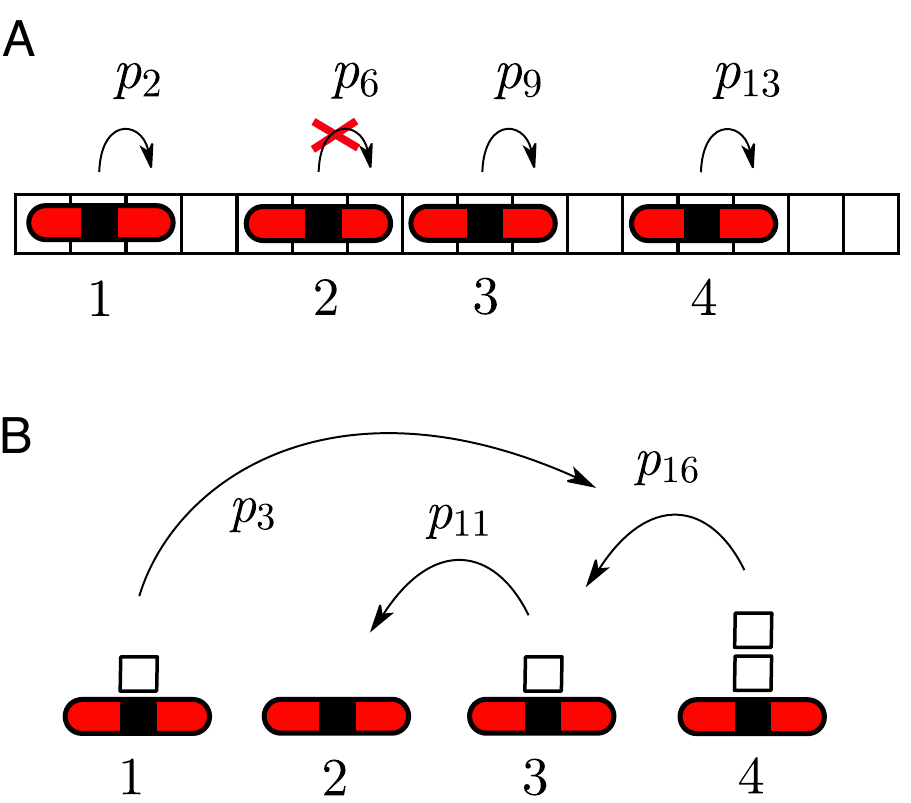}
\caption{(\textit{Related to \fref{fig:illustration}A,B and
        \sref{sec:methods_hydro_limit} of STAR Methods}). \textbf{Correspondence between inhomogeneous 
        $\ell$-TASEP (A) and the ZRP (B).} 
        $\ell$-TASEP particles (rods) correspond to ZRP sites, and holes (empty squares)
        become ZRP particles.}
        \label{suppfig:schematic_ZRP}
\end{figure}

\clearpage
\begin{figure}[p]
    \centering
    \includegraphics[width=\textwidth]{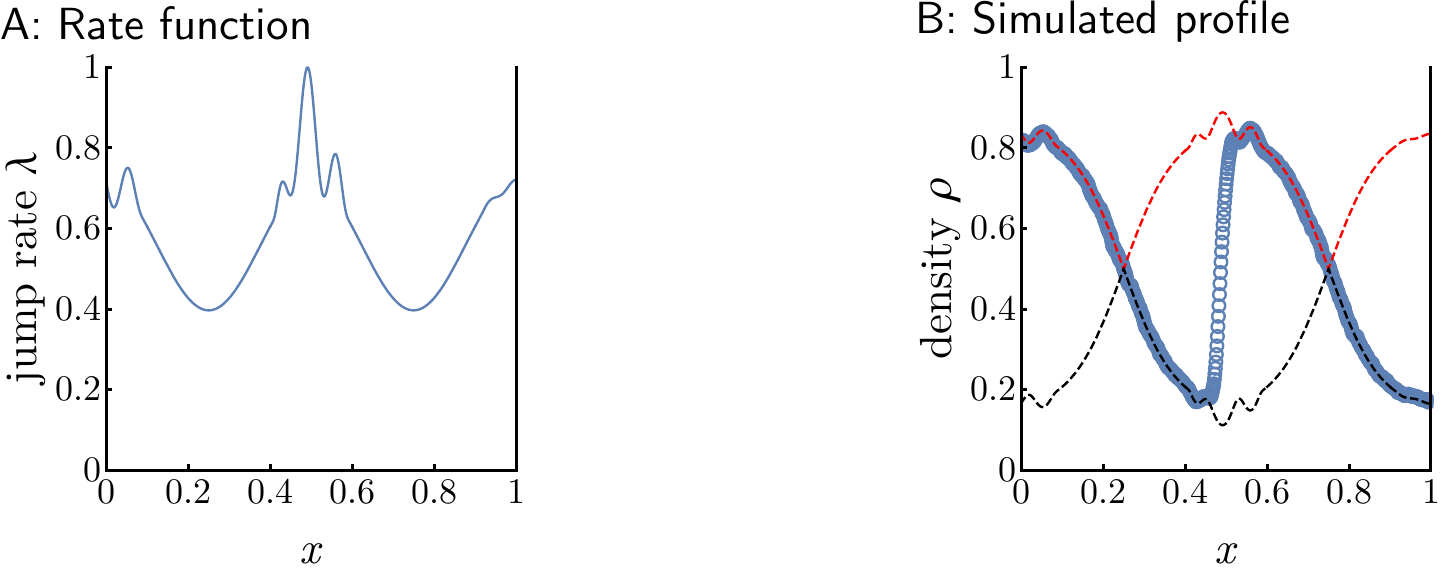}
    \caption{(\textit{Related to \sref{sec:particle_densities} and
        \sref{sec:phase_diagram_analysis} of STAR Methods}). 
        \textbf{Atypical behaviour of MC branch switching in the presence of two
        global minima.} Hydrodynamic predictions suggest that branch switching is
        bound to occur between any two global minima, but do not provide
        explicit information about the precise location of these singularities.
        Simulations indicate that branch switching is preferentially situated
        around local maxima. 
        \textbf{A:} Elongation rates. 
        \textbf{B:} Circles are averaged counts over $5\times 10^7$
        Monte-Carlo steps after $10^7$ burn-in cycles on a lattice of size $N=2000$ with
        parameters $\alpha = \beta= \ell=1$ and elongation rate function shown in \textbf{A}. We
        compare these simulated densities to the theoretical profile obtained from the 
        upper (red) and lower (black) branch solutions 
        (described in \eqref{eq:stationary_solutinos}).}
    \label{suppfig:double_defect}
\end{figure}

\begin{figure}[p]
    \centering
    \includegraphics[width=\linewidth]{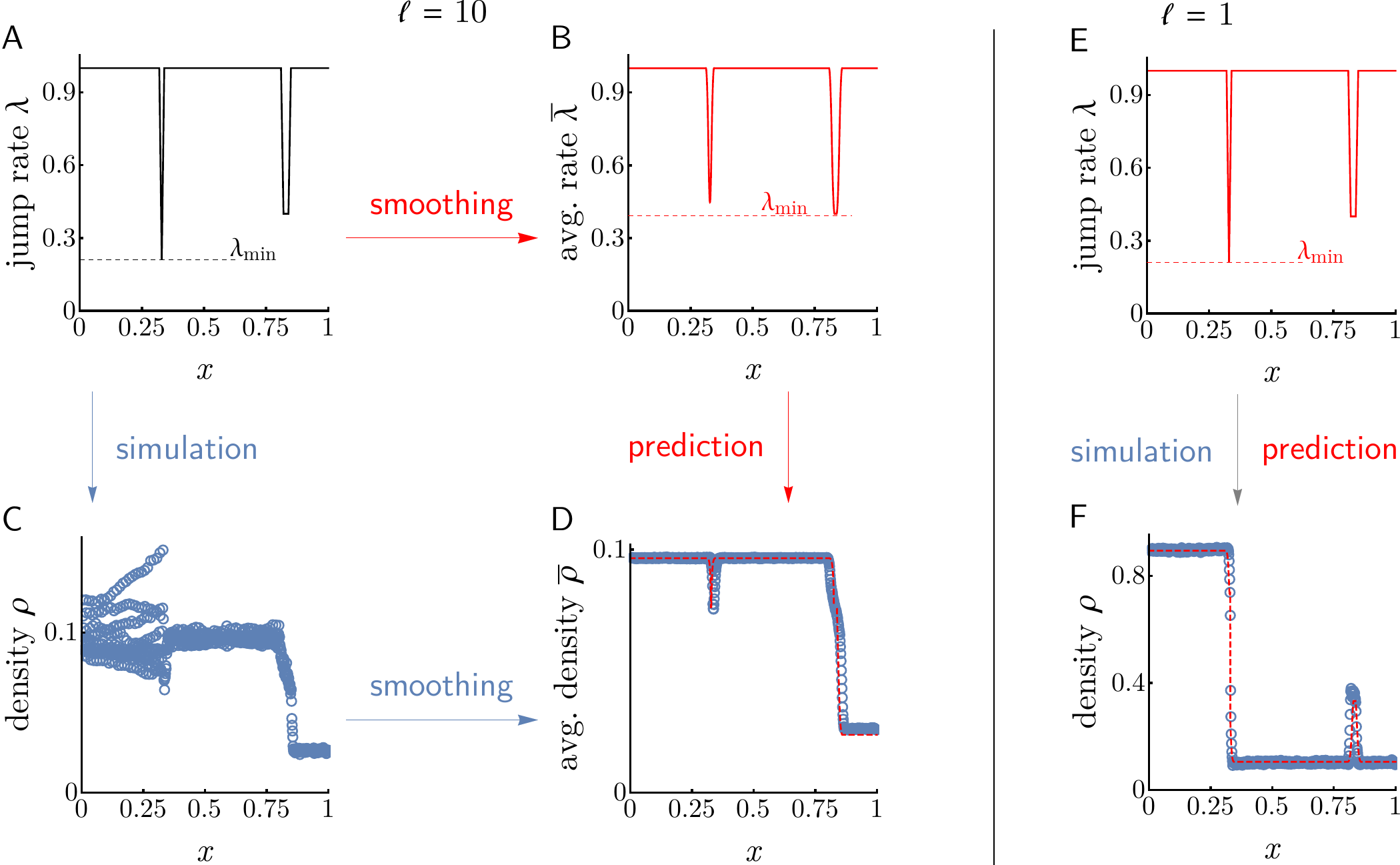}
    \caption{(\textit{Related to \fref{fig:smoothing} and
        \sref{sec:phase_transition} of STAR Methods}). 
        \textbf{MC branch switching is determined by locally averaged
        elongation rates rather than raw elongation rates (A-D), and the averaging scale 
        depends on the particle size (E,F).} Both the value as
        well as the location of the minimal elongation rate may differ
        significantly when measured with respect to the discrete
        elongation profile (Panel \textbf{A}) or smoothed elongation profile
        (Panel \textbf{B}). Panels \textbf{C} and \textbf{D} demonstrate
         that our hydrodynamic prediction is very accurate, and show that MC branch 
        switching is governed by the smoothed elongation profile rather than its discrete
        counterpart.
         Several of the yeast transcripts we analyzed are affected
        by this phenomenon, suggesting that a codon's local neighborhood is
        a stronger determinant of translation dynamics than the absolute elongation
        rate at that site.  Whether smoothed elongation rates (as opposed to unsmoothed rates) describe
        the translation dynamics more accurately is strongly linked to the particle size $(\ell)$
        and the long-range correlations (in particular, $\ell$-periodicity after
        traffic jams) it induces.   To demonstrate this point, we performed the same
        analysis as in Panels \textbf{A}-\textbf{D} using particles of size $\ell = 1$.
        We found that our hydrodynamic predictions based on the raw, unsmoothed
        elongation rates (Panel \textbf{E}) does indeed provide an accurate
        approximation of simulated densities (Panel \textbf{F}).  In short,
        the fact that the ribosome occupies 10 codons (i.e. the ``particle''
        size is $\ell = 10$) provides another reason
        (in addition to alleviating the irregularity of elongation rates that cause analytical difficulty)
        for why smoothing the elongation rates
        is the right thing to do when applying the hydrodynamic limit of the $\ell$-TASEP 
        to study mRNA translation.
    } 
        \label{suppfig:smoothing_x_min}
\end{figure}
\clearpage
\begin{figure}[p]
    \centering
    \includegraphics[width=\linewidth]{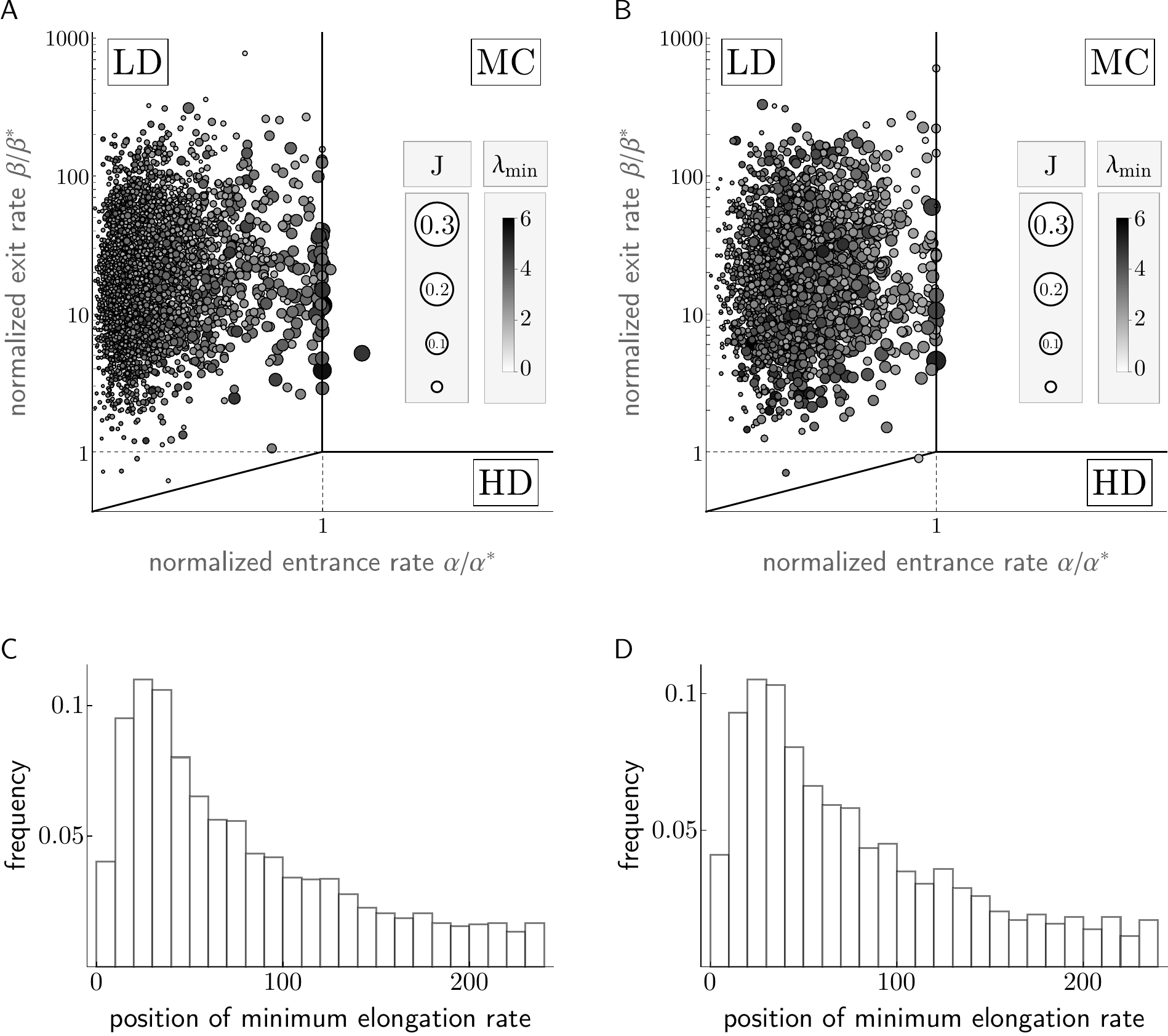}
    \caption{\textit{(Related to \sref{sec:yeast} and \fref{fig:data}).} 
    \textbf{Inferences on the efficiency of yeast's translational system are
    consistent across datasets.}
    To test the replicability of our analysis using the previously inferred
    elongation rates in \citet{daoduc2018impact} and to exclude any possible
    systematic biases, we repeated our inference on elongation rates obtained 
    by inverting \eqref{eq:first_order} on two independent ribosome profiling
    datasets: One compiled by  \citet{williams2014targeting}
    (\textbf{A, C}, total of $3098$ genes), and one by  
    \citet{pop2014} (\textbf{B, D}, total of $2536$ genes). The clear 
    localization of genes within LD and at the LD/MC boundary, together with a
    characteristic ramp-shaped distribution of the minimum elongation location
    remain apparent, lending support to our proposed design principles holding
    true not only on the 850 genes analyzed in the main manuscript, but more
    generally as a framework governing translation efficiency.}
    \label{suppfig:consistency}
\end{figure}

\clearpage
\begin{figure}[p]
    \centering
    \includegraphics[width=.85\textwidth]{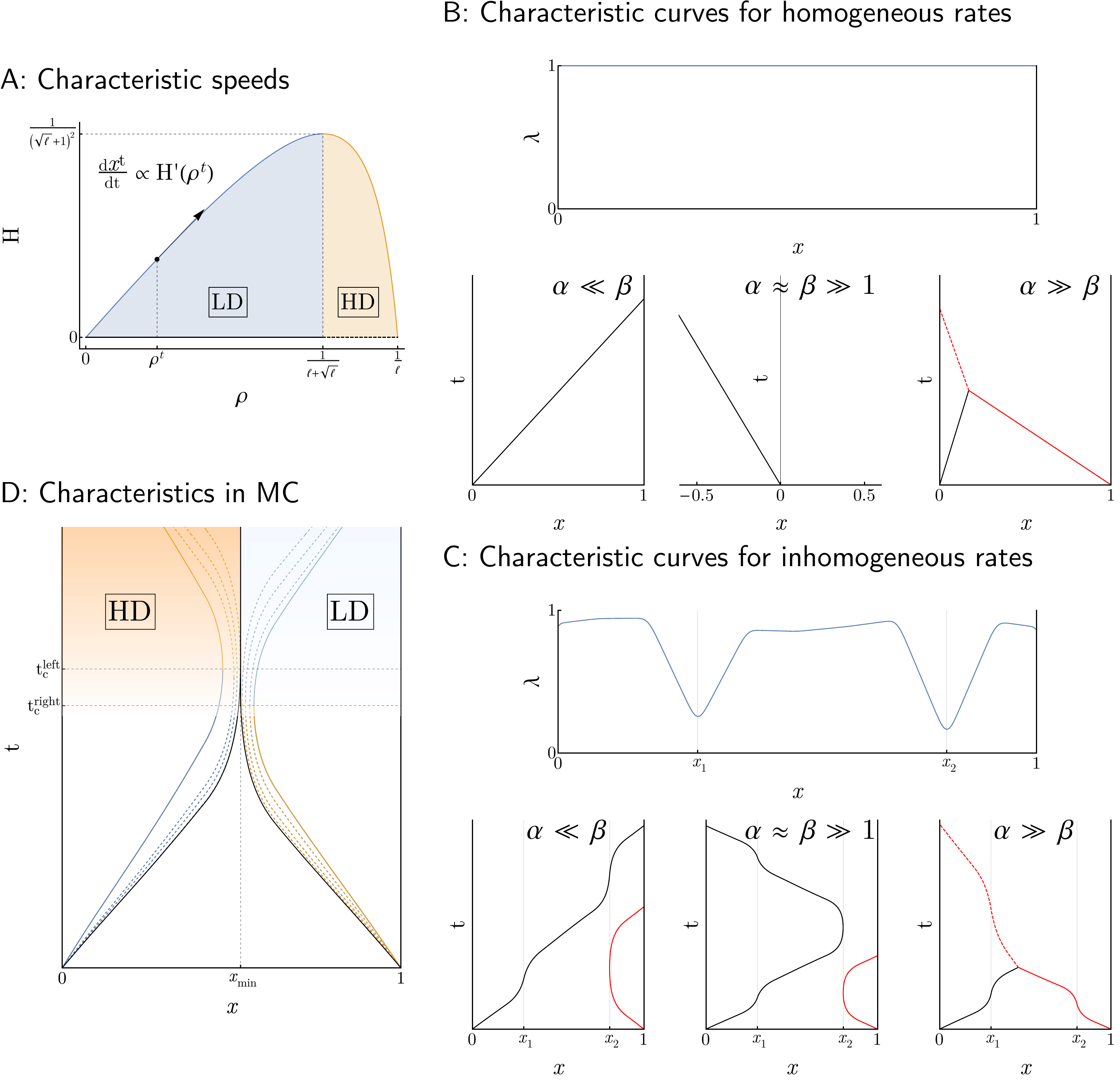}
    \caption{(\textit{Related to \sref{sec:phase_transition} of STAR Methods}).
    \textbf{$H(\rho)$ and its effect on characteristic curves.} \textbf{A:} 
    The rate-normalized flux $H(\rho) = J(\rho,x)/\lambda(x)$ is depicted in
    solid blue and orange, 
    with characteristic velocity of $x^t$ indicated. If $J(\rho^0,x^0) <
    J_{\max}$, the characteristic density $\rho^t$ stays within the regions
    marked LD (blue) or HD (orange), depending on the sign of $\rho^0 - (\ell +
    \sqrt{\ell})^{-1}$. Otherwise, $\rho^t$ may cross $(\ell + \sqrt{\ell})^{-1}$
    forcing the characteristic $x^t$ to return to its origin $x^0$. \textbf{B,C:} 
    Characteristic curves starting at lattice start (black solid curves) and end 
    (red solid curves) for different regions of the phase diagram. Dotted curves represent 
    shock fronts, with colors indicating which characteristic drives the shock.
    \textbf{B:} Homogeneous rates give rise to straight line characteristics with speed 
    $\partial_{\rho} J(\rho_0)$ and $\partial_{\rho} J(\rho_1)$, respectively.
    \textbf{C:} Inhomogeneous rates produce more complicated behavior, with 
    characteristics curves slowing down (and potentially reversing direction) near 
    the troughs ($x_1$ and $x_2$) of $\lambda$. \textbf{D:} If $J(\rho^0,x^0) > J_{\max}$ 
    the characteristic curves $x^t$ (left and right colored solid curves) reverse 
    directions at critical times $t_c$ and return to their origin. At $t_c$, the density 
    $\rho^t$ switches from LD (blue) to HD (orange) (if associated with $x^0 = 0$) or HD 
    to LD (if associated with $x^0 = 1$, cf. \textbf{A}). The same happens on all 
    associated rarefaction waves (dashed curves), which interpolate between $x^t$ and the 
    stationary shock of $x_{\max}^t$ (solid black curve).
}
    \label{suppfig:current_Diagram}
\end{figure}

\clearpage
\begin{figure}[p]
    \centering
    \includegraphics[width=0.99\linewidth]{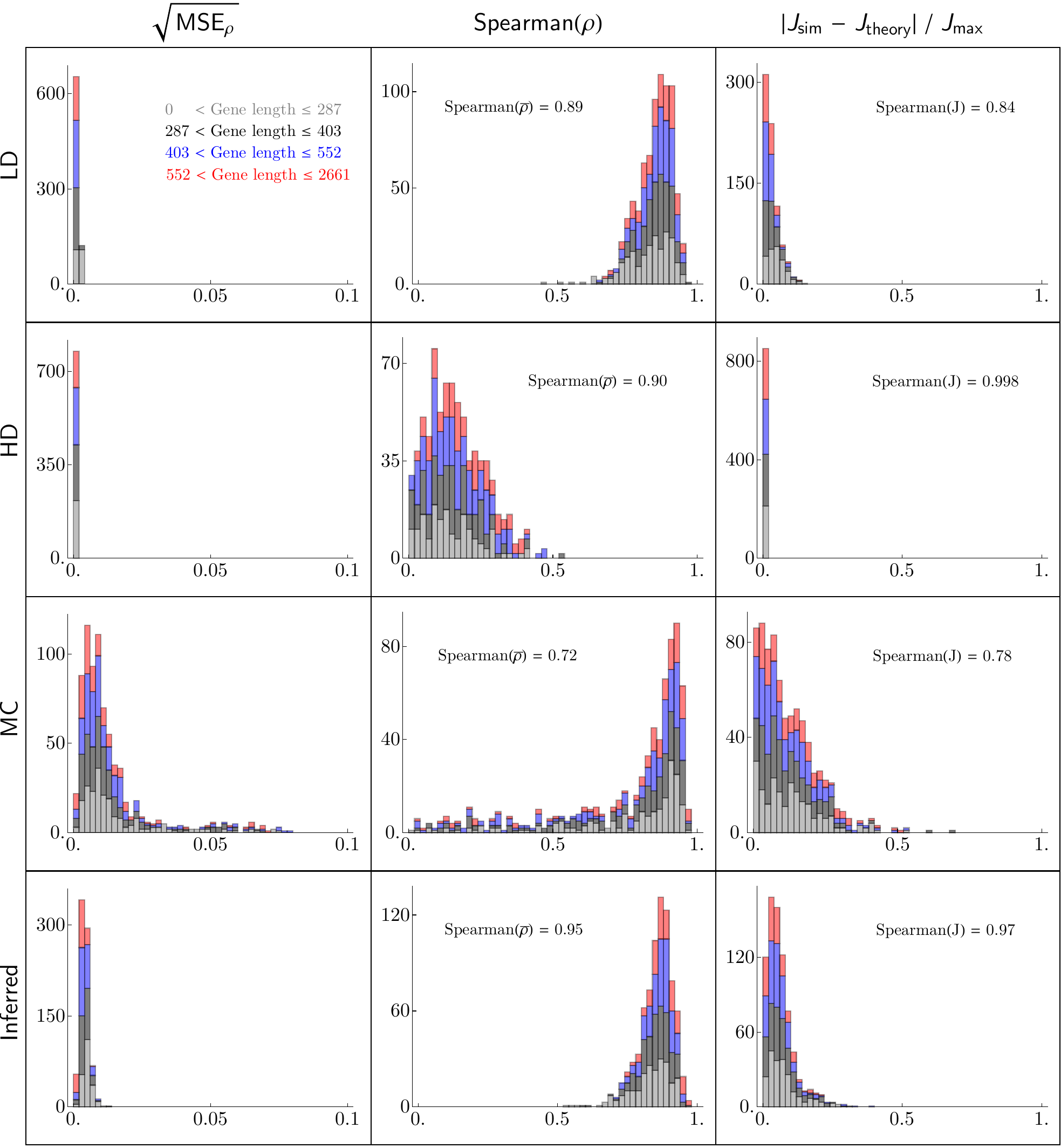}
    \caption{\textit{(Related to \sref{sec:correlations} of STAR Methods).} 
    \textbf{Comparison between simulation and theoretical prediction of our hydrodynamic approximation.}
    Absolute errors in position-specific ribosome densities $\rho$ (first column) and
    currents $J$ (third column) are low for all gene lengths (colored), for different regimes in the phase diagram (first three rows), and for biologically relevant initiation and termination
    rates (last row) inferred in \citet{daoduc2018impact}.  Moreover, transcript-by-transcript ribosome density profiles $\rho$ and mean ribosome occupancies $\overline{\rho}$ correlate well between simulated data and our hydrodynamic predictions (middle column), as does currents (figure inset in third column).
}
    \label{suppfig:correlations}
\end{figure}

\clearpage

\begin{table}[p]
    \centering
    \caption{(\textit{Related to \sref{sec:boundary_conditions} of STAR Methods}). 
            \textbf{Boundary conditions by phase.} Expected densities at the
    left ($x=0$) and right ($x=1$) end boundaries of the lattice.}
    \label{table1}
    \begin{tabularx}{\textwidth}{XXXX}
        Phase & $\rho_0$ & $\rho_1^+$ & $ \rho_1^-$ \\ \hline\\[-0.9em]
        LD  & $\dfrac{\alpha}{\lambda_0 + (\ell - 1)\alpha}$ &  
        $\dfrac{1}{\beta} \bigg[ \dfrac{\alpha(\lambda_0-\alpha)}{\lambda_0 
        + (\ell-1)\alpha} \bigg]$ &  
        $\dfrac{1}{\lambda_1} \bigg[\dfrac{\alpha(\lambda_0-\alpha)}{\lambda_0 
        + (\ell-1)\alpha} \bigg]$ \\[4mm] 
        HD & $\dfrac{1}{\ell} - \dfrac{1}{\ell\alpha}
        \bigg[\dfrac{\beta(\lambda_1-\beta)}{\lambda_1 + (\ell-1)\beta}\bigg]$ & 
        $\dfrac{\lambda_1-\beta}{\lambda_1+(\ell-1)\beta}$ & 
        $\dfrac{1}{\lambda_1} \bigg[\dfrac{\beta(\lambda_1-\beta)}{\lambda_1 + 
        (\ell-1)\beta}\bigg]$  \\[4mm]
        MC & $\dfrac{1}{\ell} -
        \dfrac{1}{\ell\alpha} \bigg[\dfrac{\lambda_{\min}}{(1+\sqrt{\ell})^2} \bigg]$& 
        $ \dfrac{1}{\beta} \bigg[\dfrac{\lambda_{\min}}{(1+\sqrt{\ell})^2}\bigg]$ & 
        $\dfrac{1}{\lambda_1}\bigg[\dfrac{\lambda_{\min}}{(1+\sqrt{\ell})^2}\bigg]$ \\[4mm]  \hline\\
    \end{tabularx}
\end{table}
\end{document}